\documentclass[12pt]{article}
\usepackage{amssymb}

\usepackage{amscd}
\usepackage{epsfig}

\newcommand{\binom}[2]{{#1 \choose #2}}

\newtheorem{theorem}{Theorem}[section]
\newtheorem{corollary}[theorem]{Corollary}

\newtheorem{lemma}[theorem]{Lemma}
\newtheorem{proposition}[theorem]{Proposition}

\newtheorem{definition}{Definition}[section]

\input{tcilatex}

\begin{document}

\begin{center}
{\Large Typical random 3-SAT formulae and the satisfiability threshold%
%TCIMACRO{
%\TeXButton{note}{\footnote[1]{\textnormal{A preliminary short version of this paper apperared in the Proceedings of the Eleventh ACM-SIAM Symposium on Discrete Algorithms, pages 124-126, San Francisco, California, January 2000.}}}}
%BeginExpansion
\footnote[1]{\textnormal{A preliminary short version of this paper apperared in the Proceedings of the Eleventh ACM-SIAM Symposium on Discrete Algorithms, pages 124-126, San Francisco, California, January 2000.}}%
%EndExpansion
}

O. Dubois$^{\dagger}$, Y. Boufkhad$^{\ddagger}$ and J. Mandler$^{\dagger}$

$^\dagger${\small LIP6, Box 169, CNRS-Universit\'{e} Paris 6, 4 place
Jussieu, 75252 Paris cedex 05, France.} \\[0pt]
{\small emails : Olivier.Dubois@lip6.fr, Jacques.Mandler@lip6.fr}

$^\ddagger${\small LIAFA, Universit\'{e} Paris 7 Denis Diderot, 175 rue du
Chevaleret, 75013 Paris, France \\[0pt]
email : boufkhad@liafa.jussieu.fr}
\end{center}

{\footnotesize \textbf{Abstract} : We present a new \emph{structural} (or
\emph{syntactic}) approach for estimating the satisfiability threshold of
random 3-SAT formulae. We show its efficiency in obtaining a jump from the
previous upper bounds, lowering them to 4.506. The method combines well with
other techniques, and also applies to other problems, such as the
3-colourability of random graphs.}

\section{Introduction\label{intro}}

The last decade has seen a growth of interest in phase transition phenomena
in hard combinatorial decision problems, due to resulting insights into
their computational complexity and that of the associated optimization
problems. There is a fast growing body of theoretical investigations as well
as ones exploring algorithmic solver implications. Latterly, moreover,
statistical physics studies have also shed new light on these phenomena,
whence a further surge in interest. Among the various and extensive
contributions, let us single out a few: \cite{ChvRee92, MitSel92,
BroFriUpf93, GenWal94, Mer98, KirKra98, Frie99, CreDau99, MonZec00,
BolBor01, DubDeq01, CocMon01, Zec201}. Several surveys can be found in \cite
{DubMonSelZec01}.\newline
One of the most challenging phase transitions, with a long history of
results, concerns the problem of 3-Satisfiability (to satisfy sets of
clauses of length $3$, i.e. disjunctions of $3$ literals). \cite
{DubMonSelZec01} contains a survey which we briefly summarize and update
here. Experiments strongly suggest that satisfiability of random 3-SAT
formulae (the 3-SAT problem) exhibits a sharp threshold or a phase
transition as a function of a \textit{control parameter}, the ratio $c$ of
the number of clauses to the number of variables. More precisely, this would
mean the existence of a critical value $c_{0}$ such that for any $c<c_{0}$
the probability of satisfiability of a random 3-SAT\ formula tends to $1$ as
$n\rightarrow \infty $, and for $c>c_{0}$ it tends to $0$. Over the years,
two series of bounds for $c_{0}$ have been established, the lower bounds
being : 2.9 (positive probability only), 2/3, 1.63, 3.003, 3.145, 3.26, 3.42
(see \cite{ChaFra90, ChvRee92, BroFriUpf93, FriSue96, Ach00, AchSor00,
KapKirLal02}), and the upper bounds: 5.191, 5.081, 4.762, 4.643, 4.602,
4.596, 4.571, 4.506 (see \cite{FraPau83, MafVeg95, KamMot95, DubBou97,
KirKra98, JanSta00, JanSta01, KapKirSta01, DubBou00}). The last bound,
4.506, was briefly presented in \cite{DubBou00}. The present paper gives a
detailed proof, emphasizing the potential of the main innovation, which we
called the \emph{structural} or \emph{syntactic} approach, in contrast to
the \emph{semantic} approach hitherto used to establish upper bounds. A few
general comments are in order. Thanks to this structural approach, a jump
from 4.643 to 4.506 was obtained. Developments since then have confirmed the
interest and versatility of this technique. Further refinements of the
semantic approach, together with subtle and sophisticated probabilistic and
analytical results, have failed to match the 4.506 bound, giving 4.571 as
announced in \cite{KapKirSta01}. And we recently applied our structural
approach to the equally challenging 3-colouring problem.\ It turned out to
combine well with the decimation technique we had used for the 3-XORSAT\
problem \cite{DubMan02}, lowering the best upper bound from 2.4945 (\cite
{FouMcD02} and references therein) to 2.427 \cite{DubMan02a}.\newline
In the ramainder of this section, we present the probabilistic model for
3-SAT we work with, then give an overview of our approach leading to the
bound of 4.506. The subsequent sections contain the detailed calculations.

\subsection{Probabilistic model. \label{Model}}

Let $V_{n}=\left\{ x_{1},...,x_{n}\right\} $ be a set of $n$ \textit{boolean
variables, }$L_{n}=\left\{ x_{1},\bar{x}_{1},...,x_{n},\bar{x}_{n}\right\} $
the corresponding set of positively and negatively signed \textit{literals}.
In this paper we use the \textit{ordered-clauses} model. Here an $n$-formula
$F$ is simply a map to $L_{n}$ from the \textit{formula template }$\Lambda
_{c,n}$, an array of $cn$ \textit{clause templates }consisting each of $3$\
ordered\textit{\ places} or \textit{cells}. If the literal $l$ is the image
under $F$ of cell $\xi $, we also say that it \textit{fills }$\xi $. The set
$\Omega \left( n,c\right) $ of $n$-formulae is made into a probability space
by assigning each formula the probability $1/\left| \Omega \left( n,c\right)
\right| =\left( 2n\right) ^{-3cn}$.

Each \textit{truth assignment }$\mathcal{A}:V_{n}\rightarrow \left\{
0,1\right\} $ is conventionally extended to $L_{n}$ so that $\mathcal{A}%
\left( \bar{x}_{i}\right) =1-\mathcal{A}\left( x_{i}\right) $, and is said
to \textit{satisfy} the clause $C_{k}$ if $\mathcal{A}\left( l\right) =1$
for some $l\in C_{k}$, and the formula $F$ if it satisfies all its clauses;
in which case $\mathcal{A}$ is a \textit{solution }of $F$, and $F$ is
\textit{satisfiable}. The probability of satisfiability of a random formula $%
F$ of $\Omega \left( n,c\right) $ is denoted by $\mathbf{Pr}_{n,c}\left(
SAT\right) .$

A few words in comparison with the \textit{non-ordered-clauses} model, also
very usual.\ Here a clause is a \textit{set} of 3\ literals with \textit{%
distinct} underlying variables, and a random formula is a \textit{sequence}
of $m=cn$ clauses drawn independently and uniformly among the $2^{3}{{{{{{{{{%
{{{{{\binom{n }{3}}}}}}}}}}}}}}}$ possible clauses. Convergence to $0$
(resp. $1$), as $n\rightarrow \infty ,$ of $\mathbf{Pr}_{n,c}\left(
SAT\right) $ is readily seen to imply the same for the probability in the
non-ordered-clauses model. Thus our upper bound of $4.506$, once proven in
the ordered-clauses model, will hold in both.

\subsection{Outline.\label{outline}}

We give first a general idea of our approach stemming from concrete
experiments. A computer-based generator of random formulae churns out
\textit{mechanically}, as the case may be, \textit{only} satisfiable, or
\textit{only} contradictory formulae.\ To say that certain formulae are
\textit{never} produced (within a realistic timeframe) simply means that
they form a set of vanishingly small probability; and, due to the very
dumbness of the generator, the distinction between `likely' and `unlikely'
formulae must be possible on a very basic level, considering only their
\textit{form} or \textit{structure}. Ideally, we would like an exact
criterion for `likely' or `typical' formulae; possibly, then, the first
moment method, restricted not to particular types of \textit{solutions}, but
to \textit{formulae} with this particular property, might give us the exact
value of $c_{0}.$ Such an exact characterization is elusive, though, and
unlikely to emerge in a simple, usable form. Rather, in this paper we show
the usefulness of an uncomplicated partial characterization in terms of the
numbers of occurrences and signs of the variables. The pure effect on the
expectation of restricting the formulae becomes only part of the story.
Equally important is the fact that, far from interfering with other
approaches, the added structure actually helps in otherwise difficult or
hopeless enumerations. Thus we do not need, e.g., sophisticated probability
results.\ And, particularly, we are able to introduce at virtually no cost
some structural manipulations on the balancing of the signs of occurrences
per variable which would be impractical in the purely semantic approach. On
the other hand, to attain full rigour the method does require fairly lengthy
calculations, notably to bound errors arising from the finite size of
formulae, and thoroughly to justify the optimization procedures. These
remain relatively elementary, though, and, in the case of the error
estimates, fairly routine.

Practically we first characterize the asymptotic distribution of the signed
occurrences per variable, namely :

\begin{lemma}
\label{typstruct} For any integers $0\leq p\leq x$, define $\kappa
_{x,p}=2^{-x}{{\binom{x}{p}}}p\left( x,\lambda \right) $, where $\lambda =3c$
and $p\left( x,\lambda \right) $ is the Poisson probability mass function of
mean $\lambda $, i.e. $p\left( x,\lambda \right) =e^{-\lambda }\lambda
^{x}/x!$. Let the random variable $\omega _{x,p}$ be the proportion of
variables of a random formula having $x$ occurrences, among which exactly $p$
have a positive signature. Then for any $\varepsilon >0$, $%
\lim_{n\rightarrow \infty }\mathbf{\Pr }\left( \left| \omega _{x,p}-\kappa
_{x,p}\right| >\varepsilon \right) =0$.
\end{lemma}

It can be seen easily that Lemma \ref{typstruct} implies that an upper bound
on the satisfiability threshold is obtained by calculating the expected
number of solutions of a \textit{typical} formula (in a sense to be
specified shortly, but roughly meaning that for most $(x,p)$, there are
nearly $\kappa _{x,p}n$ variables having $x$ total, $p$ positive,
occurrences). Typical formulae, however, also provide us a strong means to
go further in the structural manipulation of formulae. But we need first to
recall the definition of particular solutions which in \cite{DubBou97} we
called PPSs (for Positively Prime Solution, symmetrically there are NPSs).
Note that these restrictive solutions have been introduced independently by
Kirousis\emph{\ et al}. in \cite{KirKra98} under the terminology of locally
maximal solutions and single-flip technique.

\begin{definition}
\label{defPPS}A Positively Prime Solution (PPS) $\mathcal{A}$ of a SAT
formula $F$ is a solution of $F$ such that no variable of $F$ with the value
$1$ under $\mathcal{A}$ can be singly inverted (or switched) to $0$ unless
at least one of the clauses of $F$ becomes unsatisfied, that is, the new
assignment is no longer a solution of $F.$
\end{definition}

Any satisfiable formula has a PPS, but some have very many: they provide
extremely useful, yet somewhat limited restriction. A means to enhance this
is \textit{unbalancing}, which we now introduce on an intuitive level.

When enumerating formulae with a view to computing an expectation, we
usually count as different some formulae (very many, in fact) which really
are the same from the point of view of satisfiability. This happens in more
than one way. Some formulae differ from each other by a permutation on the
set of clauses, or on the set of variables; these, however, are fairly
transparent.\ What concerns us here are formulae deduced from one another by
\textit{renaming }certain variables, in the restricted (and usual) sense of
inverting the signs of all their occurrences. Their significance to us stems
from the fact that unlike those just mentioned, they are not neutral with
respect to PPSs. Consider, e.g., a \textit{pure} variable (one which has all
of its occurrences of the same sign).\ This sign is indifferent as far as
ordinary solutions are concerned, but a solution in which a negated pure
variable takes the value $1$ \textit{cannot} be a PPS, while an unnegated
pure variable has the best chances that many of the solutions giving it the
value $1$ will be PPSs. Similarly, a variable with more positive than
negative instances is likely to kill fewer PPSs than the reverse. Therefore,
of two formulae which differ only by the systematic inversion of some
variables, the one with more negatively unbalanced variables may be assumed
to have fewer PPSs. To be precise, call two formulae \textit{equivalent} if
one can be obtained from the other by renaming certain variables. Clearly,
this is indeed an equivalence relation $\mathcal{R}$ on the set $\Omega
(n,c) $ of 3-SAT formulae on $n$ variables with $cn$ clauses; $\mathcal{R}$
results in the partitioning of formulae with respect to equivalence modulo
variable renaming, and the cardinality of the equivalence class of a formula
$F$ is \thinspace $2^{v_{u}(F)}$, where $v_{u}(F)$ is the number of \textit{%
unbalanced} variables in $F$ (variables having unequal numbers of positive
and negative occurrences; note that an absent variable is, by definition,
balanced).

Since negatively unbalanced variables tend to inhibit PPSs, we have a good
candidate for the formula with the fewest PPSs within each equivalence class
$\mathcal{C}$: namely, the \textit{totally unbalanced representative }$F^{-}$
obtained from any $F\in \mathcal{C}$ by renaming exactly those variables
which have more positive than negative occurrences. Moreover we have an easy
criterion for a formula to be the totally unbalanced representative of a
typical formula, namely that the proportion of variables having $x$ total, $%
p $ positive occurrences be $2\kappa _{x,p}$ if $x>p$; $\kappa _{x,p}$ if $%
x=2p $; and $0$ if $x<2p$. So these representatives, or, as we shall say,
the typical totally unbalanced formulae (by abuse of language, since they
are actually not typical at all) can be defined just like the typical
formulae, only using instead of the $\kappa _{x,p}$'s their totally
unbalanced counterparts, the $\widetilde{\kappa }_{x,p}$'s defined by :
\begin{equation}
\widetilde{\kappa }_{x,p}=\left\{
\begin{array}{lll}
2\kappa _{x,p} & \mathrm{if} & x>2p \\
\kappa _{x,p} & \mathrm{if} & x=2p \\
0 & \mathrm{if} & x<2p
\end{array}
\right.  \label{ktilda}
\end{equation}
All equivalence classes of such representatives have the same number of
elements, namely: $2^{2n\sum_{x>2p}\kappa _{x,p}}=2^{n\sum_{x>2p}\widetilde{%
\kappa }_{x,p}}$.

Calculations with typical totally unbalanced formulae are no harder than
with plain and ordinary typical formulae, in fact they are much the same
with $\widetilde{\mathbf{\kappa }}$ replacing $\mathbf{\kappa },$ and the
specifics of the distribution tend to intervene only in the very last
stages. Computing (\textit{via} a simple technical device) the expected
number of PPSs of the former rather than the latter, then multiplying by the
above size of equivalence clases, we get what amounts to a `skewed'
expectation where each formula is counted, not according to its own number
of PPSs, but to that of its representative with fewest PPSs. It is this,
combined with the gain already inherent in the restriction to structured
formulae \textit{per se}, that affords us a very significant improvement on
the upper bound of 4.643 resulting from the expectation of PPSs alone \cite
{DubBou97}.

Before proceeding, we have to take account of some practical remarks raised
by the foregoing considerations. (i) The $\kappa _{x,p}$'s or the $%
\widetilde{\kappa }_{x,p}$'s constitute an infinite family and are all $\neq
0$, while a formula has finite length; (ii) The proportions $\nu _{x,p}$ of
variables having $x$ total, $p$ positive occurrences in a formula $F\in
\Omega \left( c,n\right) $ must verify $\sum \nu _{x,p}=1$ and $\sum x\nu
_{x,p}=3c=\lambda $, where the sums are in effect finite, while the
equalities $\sum \kappa _{x,p}=1$ and $\sum x\kappa _{x,p}=\lambda $ only
apply with infinite sums (series); (iii) The $\kappa _{x,p}$'s are
irrational, so they cannot be exact proportions even for special values of $%
n $. Thus, in order to derive a rigorous argument, we define what we call
formulae obeying a given distribution of signed occurrences to a specified
approximation :

\begin{definition}
\label{defob} Let $\Xi =(\xi _{x,p})_{0\leq p\leq x}$ be a family of
nonnegative real numbers satisfiying the relations $\sum_{x=0}^{\infty
}\sum_{p=0}^{x}\xi _{x,p}=1$ and $\sum_{x=0}^{\infty }\sum_{p=0}^{x}x\xi
_{x,p}=\lambda .$ Given a real $\varepsilon >0$ and an integer $x_{max}$, a
formula $F\in \Omega (n,c)$ is said to \textit{obey the distribution} $\Xi $
\textit{to the accuracy} $(\varepsilon ,x_{max})$ iff for $0\leq p\leq x\leq
x_{max}$, the number of variables having $x$ occurrences in $F$, $p$ of
which are positive, lies between $(\xi _{x,p}-\varepsilon )n$ and $(\xi
_{x,p}+\varepsilon )n$. The set of formulae in $\Omega (n,c)$ obeying $\Xi $
to the accuracy $(\varepsilon ,x_{max})$ will be denoted by $\mathcal{F}(\Xi
,\varepsilon ,x_{max},n,c)$.
\end{definition}

The term `typical formula' will sometimes be used loosely to indicate a
formula which obeys the distribution $(\kappa _{x,p})$ to the accuracy $%
(\varepsilon ,x_{max})$ for some (large) $x_{max}$ and some (small) $%
\varepsilon $.

Henceforth the distributions of the $\kappa _{x,p}$'s and of the $\widetilde{%
\kappa }_{x,p}$'s (corresponding, of course, to some value of $\lambda =3c$)
will be denoted by $\Xi _{0}$ and $\widetilde{\Xi }_{0}$, respectively.
Also, when the context makes the various parameters clear, we will often use
the abbreviated notation $\mathbf{E}[PPS]$ for the expected number of PPSs
of formulae drawn uniformly from $\mathcal{F}(\Xi ,\varepsilon ,x_{max},n,c)$%
. Strictly speaking, a direct calculation of the expectation of PPSs of
typical totally unbalanced formulae would involve an awkward change of
probability space. The same end result can be achieved much more
conveniently by introducing an \textit{ad hoc} r.v.\ on the original
probability space $\Omega (n,c)$, then linking its expectation to the
probability of satisfiability:

\begin{proposition}
\label{domtyp} Define the r.v. $X_{n,\varepsilon ,x_{max},c}$ on $\Omega
(n,c)$ by:
\[
X_{n,\varepsilon ,x_{max},c}(F)=\left\{
\begin{array}{l}
\begin{array}{lll}
2^{n\sum_{x>2p}\widetilde{\kappa }_{x,p}}\times PPS(F) & \mathrm{if} & F\in
\mathcal{F}(\widetilde{\Xi }_{0},\varepsilon ,x_{max},n,c)
\end{array}
\\
\begin{array}{ll}
0 & \mathrm{otherwise}
\end{array}
\end{array}
\right.
\]
and set $\rho =\rho _{x_{max}}=\sum_{2p>x_{max}}\kappa _{2p,p},\;\;\;\Delta
=\Delta _{x_{max}}=1/2\;\left( x_{max}/2+1\right) .$ If, for some integer $%
x_{max}$ and some $\varepsilon >0$, $2^{(\rho +\varepsilon \Delta )n}.%
\mathbf{E}[X_{n,\varepsilon ,x_{max},c}]$ tends to $0$ as $n\rightarrow
\infty $, then so does $\mathbf{Pr}_{n,c}\left( SAT\right) $.
\end{proposition}

(\textbf{Remark:} It will be clear from the proof that this remains true if
instead of PPSs we use any class of solutions such that any satisfiable
formula possesses at least one solution in this class, e.g. prime implicants
\cite{BouDub99}, `double flips' \cite{KirKra98}.)

The rest of our plan will be to compute an explicit expression of $\mathbf{E}%
[X_{n,\varepsilon ,x_{max},c}]$ as sums of combinatorial terms, then an
asymptotic exponential upper bound of this expectation.\ This will be
obtained as a function of values of parameters satisfying a system of
equations, which will be reduced to two equations in two unknowns. Careful
study of these equations, coupled with numerical calculations, will show
that for $\widetilde{\Xi }_{0}=(\widetilde{\kappa }_{x,p})$, $c=4.506$, and
appropriate values of $x_{max}$ and $\varepsilon $, $2^{\left( \rho
_{x_{max}}+\varepsilon \Delta _{x_{max}}\right) n}\mathbf{E}%
[X_{n,\varepsilon ,x_{max},c}]$ tends to $0$ as $n\rightarrow \infty $.

\section{Basic structural results on random 3-SAT formulae\label{basic}}

We have first to prove Lemma \ref{typstruct}. Here the classical limit
theorems of probability do not apply, and some form of large-deviation
inequality has to be used. One method is to first obtain the expectation of $%
\omega _{x,p}$ as $\kappa _{x,p},$ then apply the method of bounded
differences (see, e.g., \cite{Hab98}, pp. 16, 221). Or, a proof using
Poissonization may be of independent interest, giving stronger bounds, so we
include a detailed one in Appendix A. Interestingly, Lemma \ref{typstruct},
which is all we need, uses the full power of neither approach.

The quantity $|\left\{ \left( x,p\right) :0\leq p\leq x\leq x_{max}\right\}
|=(x_{max}+1)(x_{max}+2)/2$ is encountered repeatedly in the sequel, we
denote it $D\left( x_{max}\right) $ or simply $D.$

\textbf{Proof of Proposition \ref{domtyp}. } With the equivalence relation $%
\mathcal{R}$ as in Section \ref{outline}, and $\widehat{\mathcal{R}}$
induced by $\mathcal{R}$ on $\mathcal{F}(\Xi _{0},\frac{\varepsilon }{2}%
,x_{max},n,c)$, the quotient (canonical) map $\mathcal{F}(\Xi _{0},\frac{%
\varepsilon }{2},x_{max},n,c)\rightarrow \mathcal{F}(\Xi _{0},\frac{%
\varepsilon }{2},x_{max},n,c)/\widehat{\mathcal{R}}$ maps $F$ to the (class
of the) formula $F^{-}$ obtained by renaming all variables of $F$ having
more positive than negative occurrences.\newline
Recall that $\omega _{x,p}(F)$ denotes the proportion of variables in a
formula $F\in \Omega (n,c)$ having $x$ total, $p$ positive occurrences. Then
:
\[
\omega _{x,p}(F^{-})=\left\{
\begin{array}{lll}
\omega _{x,p}(F)+\omega _{x,x-p}(F) & \mathrm{if} & x>2p \\
\omega _{x,p}(F) & \mathrm{if} & x=2p \\
0 & \mathrm{if} & x<2p
\end{array}
\right.
\]
A single $F^{-}\in \mathcal{F}(\Xi _{0},\frac{\varepsilon }{2},x_{max},n,c)/%
\widehat{\mathcal{R}}$ may come from at most $2^{v_{u}(F^{-})}$ formulae
(not all necessarily in $\mathcal{F}(\widetilde{\Xi }_{0},\frac{\varepsilon
}{2},x_{max},n,c)$). Taking into account that if $x>2p$, we have $\widetilde{%
\kappa }_{x,p}=\kappa _{x,p}+\kappa _{x,x-p}$ (because $\kappa _{x,p}=\kappa
_{x,x-p}$), we have:
\[
|\omega _{x,p}(F^{-})-\tilde{\kappa}_{x,p}|=\left\{
\begin{array}{l}
|\omega _{x,p}(F)-\kappa _{x,p}|+|\omega _{x,x-p}(F)-\kappa _{x,x-p}|\leq
\frac{\varepsilon }{2}+\frac{\varepsilon }{2}=\varepsilon \;\;\mathrm{if\;\;}%
x>2p \\
|\omega _{x,p}(F)-\kappa _{x,p}|\leq \frac{\varepsilon }{2}<\varepsilon \;\;%
\mathrm{if\;\;}x=2p \\
0\;\;\mathrm{if\;\;}x<2p,
\end{array}
\right.
\]
so that $\left| \mathcal{F}(\Xi _{0},\frac{\varepsilon }{2},x_{max},n,c)/%
\widehat{\mathcal{R}}\right| \leq \left| \mathcal{F}(\widetilde{\Xi }%
_{0},\varepsilon ,x_{max},n,c)\right| .$ Further,
\begin{eqnarray*}
\frac{v_{u}(F^{-})}{n} &=&\sum_{0\leq 2p<x}\omega _{x,p}(F^{-})\leq
1-\sum_{0\leq 2p\leq x_{max}}\omega _{2p,p}(F^{-}) \\
&\leq &1-\sum_{0\leq 2p\leq x_{max}}\widetilde{\kappa }_{2p,p}+\left( \frac{%
x_{max}}{2}+1\right) \frac{\varepsilon }{2} \\
&=&\sum_{0\leq 2p<x}\widetilde{\kappa }_{x,p}+\sum_{2p>x_{max}}\widetilde{%
\kappa }_{2p,p}+\left( \frac{x_{max}}{2}+1\right) \frac{\varepsilon }{2}
\end{eqnarray*}
Therefore, since $\widetilde{\kappa }_{2p,p}=\kappa _{2p,p,}$%
\begin{eqnarray*}
\left| \mathcal{F}(\Xi _{0},\varepsilon ,x_{max},n,c)\right| &\leq
&2^{2n\sum_{0\leq 2p<x}\kappa _{x,p}}\times 2^{\left( \varepsilon \Delta
_{x_{max}}+\rho _{x_{max}}\right) n}\times \left| \mathcal{F}(\Xi _{0},\frac{%
\varepsilon }{2},x_{max},n,c)/\widehat{\mathcal{R}}\right| \\
&\leq &2^{2n\sum_{0\leq 2p<x}\kappa _{x,p}}\times 2^{\left( \varepsilon
\Delta _{x_{max}}+\rho _{x_{max}}\right) n}\times \left| \mathcal{F}(%
\widetilde{\Xi }_{0},\varepsilon ,x_{max},n,c)\right|
\end{eqnarray*}

\textbf{Remark.} Our bound on $v_{u}(F^{-})$ might at first sight seem too
loose, since, instead of allowing all unbalanced variables to be renamed in
any combination, we should really pick half of each group of $\widetilde{%
\kappa }_{x,p}.n$ and rename only these. Actually, the two bounds do not
differ in their exponential orders of growth as $n\rightarrow \infty $.%
\newline
Note that $F$ is satisfiable iff $F^{-}$ is. So,
\begin{eqnarray*}
\left| \mathcal{F}(\Xi _{0},\frac{\varepsilon }{2},x_{max},n,c)\cap
SAT(n,c)\right| &\leq &2^{2n\sum_{0\leq 2p<x}\kappa _{x,p}}\times
2^{\varepsilon \Delta n} \\
&&\times \left| \mathcal{F}(\widetilde{\Xi }_{0},\varepsilon
,x_{max},n,c)\cap SAT(n,c)\right|
\end{eqnarray*}
We are now able to show that if $2^{\left( \rho +\varepsilon \Delta \right)
n}\times \mathbf{E}[X_{n,\varepsilon ,x_{max},c}]$ tends to $0$, then so
does the probability of satisfiability. Indeed:
\begin{eqnarray*}
\mathbf{Pr}_{n,c}\left( SAT\right) &=&\frac{|SAT(n,c)|}{|\Omega (n,c)|} \\
&\leq &\frac{|\mathcal{F}(\Xi _{0},\frac{\varepsilon }{2},x_{max},n,c)\cap
SAT(n,c)|}{|\Omega (n,c)|} \\
&&+\sum_{0\leq p\leq x\leq x_{max}}\frac{|\{F\in SAT(n,c):\mathrm{\ }|\omega
_{x,p}(F)-\kappa _{x,p}|>\frac{\varepsilon }{2}\}|}{|\Omega (n,c)|}
\end{eqnarray*}
\linebreak By Lemma \ref{typstruct}, each of the $D\left( x_{max}\right) $
terms of the last sum tends to $0$ as $n\rightarrow \infty $, hence:
\[
\mathbf{Pr}_{n,c}\left( SAT\right) \leq \frac{2^{2n\sum_{0\leq 2p<x}\kappa
_{x,p}}\times 2^{(\varepsilon \Delta +\rho )n}\times \left| \mathcal{F}(%
\widetilde{\Xi }_{0},\varepsilon ,x_{max},n,c)\cap SAT(n,c)\right| }{|\Omega
(n,c)|}+o(1)
\]
So,
\[
\mathbf{Pr}(SAT)\leq \frac{2^{2n\sum_{0\leq 2p<x}\kappa _{x,p}}\times
2^{(\varepsilon \Delta +\rho )n}}{|\Omega (n,c)|}\times \sum_{F\in \mathcal{F%
}(\widetilde{\Xi }_{0},\varepsilon ,x_{max},n,c)\cap SAT(n,c)}1+o(1)
\]
Now, since any satisfiable formula has at least one PPS, we can write:
\begin{eqnarray*}
\mathbf{Pr}(SAT) &\leq &\frac{2^{(\varepsilon \Delta +\rho )n}}{|\Omega
(n,c)|}\times \sum_{F\in \mathcal{F}(\widetilde{\Xi }_{0},\varepsilon
,x_{max},n,c)}2^{2n\sum_{0\leq 2p<x}\kappa _{x,p}}\times PPS(F)+o(1) \\
&=&\frac{2^{(\varepsilon \Delta +\rho )n}}{|\Omega (n,c)|}\times \sum_{F\in
\mathcal{F}(\widetilde{\Xi }_{0},\varepsilon ,x_{max},n,c)}X_{n,\varepsilon
,x_{max},c}(F)+o(1)=2^{\left( \rho +\varepsilon \Delta \right) n}\times
\mathbf{E}[X_{n,\varepsilon ,x_{max},c}]+o(1)
\end{eqnarray*}
\hfill ${\blacksquare }$

\section{Combinatorial analysis of the expectation.}

\subsection{\textbf{The set }$\Theta _{\varepsilon ,x_{max},n,c}$\textbf{.%
\label{deftheta}}}

In order to estimate the expected number of PPSs of formulae in $\mathcal{F}(%
\widetilde{\Xi }_{0},\varepsilon ,x_{max},n,c)$, we shall first compute the
number of such formulae having fixed values of the proportions $\omega
_{x,p}(F)$ for $0\leq p\leq x\leq x_{max}$. It will be convenient to
characterize these formulae as associated with an element of the set $\Theta
_{\varepsilon ,x_{max},n,c}\subseteq \Bbb{Q}^{D}$ of vectors $\mathbf{\theta
}=(\theta _{x,p})_{0\leq p\leq x\leq x_{max}}$ such that (with the notation $%
I_{n}=\left\{ 0,\frac{1}{n},\frac{2}{n},...,\frac{n-1}{n},1\right\} $, which
applies throughout the sequel):
\begin{eqnarray*}
\mathrm{(i)\ \ \ \ \ \ \ \ \ \ \ \ \ \ \ \ \ \ \ \ \ \ \ \ \ \ \ \ \ }\theta
_{x,p} &\in &I_{n},\;\;0\leq p\leq x\leq x_{max}; \\
\mathrm{(ii)}\;\;\;\;\;\;\;\;\;\;\;\;\;\;\;\;\;\;\;\;\sum_{x=0}^{x_{max}}%
\sum_{p=0}^{x}\theta _{x,p} &\leq &1; \\
\mathrm{(iii)\ \ \ \ \ \ \ \ \ \ \ \ \ \ \ \ \ }\left| \theta _{x,p}-%
\widetilde{\kappa }_{x,p}\right| &\leq &\varepsilon ,\;\;0\leq p\leq x\leq
x_{max};
\end{eqnarray*}

It is clear that a formula $F$ is in $\mathcal{F}(\widetilde{\Xi }%
_{0},\varepsilon ,x_{max},n,c)$ iff the vector $(\omega _{x,p}(F))_{0\leq
p\leq x\leq x_{max}}$ is in $\Theta _{\varepsilon ,x_{max},n,c}$. For $%
\mathbf{\theta }\in \Theta _{\varepsilon ,x_{max},n,c}$, we denote by $%
\mathcal{F}(\mathbf{\theta })$ the subset of $\mathcal{F}(\widetilde{\Xi }%
_{0},\varepsilon ,x_{max},n,c)$ consisting of those formulae $F$ such that
for $0\leq p\leq x\leq x_{max}$, $\omega _{x,p}(F)=\theta _{x,p}$. We are
able to focus on the number of elements of $\mathcal{F}(\mathbf{\theta })$
mainly because, as the following lemma shows, the relatively small (i.e.
polynomial) size of $\Theta _{\varepsilon ,x_{max},n,c}$ means that, as far
as exponential orders are concerned, it makes no real difference whether $%
\mathbf{\theta }$ is kept fixed or allowed to vary within $\Theta
_{\varepsilon ,x_{max},n,c}$:

\begin{lemma}
\label{thetasize}$\left| \Theta _{\varepsilon ,x_{max},n,c}\right| \leq
\left( 2\varepsilon n\right) ^{D}.$
\end{lemma}

\begin{proof}
If the vector $\mathbf{\theta }$ is in $\Theta _{\varepsilon ,x_{max},n,c},$
then for $0\leq p\leq x\leq x_{max}$, $\theta _{x,p}n$ is an integer
comprised between $(\widetilde{\kappa }_{x,p}-\varepsilon )n$ and $(%
\widetilde{\kappa }_{x,p}+\varepsilon )n$, so there are at most $%
2\varepsilon n$ possible values for $\theta _{x,p}$.
\end{proof}

\subsection{Counting formulae with a given PPS and fixed proportions of
variables having given numbers of occurrences\label{fixtheta}}

For some given $\varepsilon $ and $x_{max}$, we now consider a fixed vector $%
\mathbf{\theta }\in \Theta _{\varepsilon ,x_{max},n,c}$ and a truth value
assignment $\mathcal{A\in }\left\{ 0,1\right\} ^{n}$, identified with the
subset $\mathcal{A}^{-1}(1)$ of the set of variables $V_{n}$. Let $\mathcal{F%
}(\mathbf{\theta },\mathcal{A})$ be the set of formulae $F\in \Omega (n,c)$
such that $\mathcal{A}$ is a PPS of $F$ and that for $0\leq p\leq x\leq
x_{max}$, $\omega _{x,p}(F)=\theta _{x,p}$. Thus:

\begin{proposition}
\label{Efromfixtheta}\textbf{\ }$\mathbf{E}\left( X_{n,\varepsilon
,x_{max},c}\right) =\frac{2^{n\sum_{x>2p}\widetilde{\kappa }_{x,p}}}{\left|
\Omega (n,c)\right| }\sum\limits_{\mathbf{\theta }\in \Theta _{\varepsilon
,x_{max},n,c}}\sum\limits_{\mathcal{A\in }\left\{ 0,1\right\} ^{n}}\left|
\mathcal{F}(\mathbf{\theta },\mathcal{A})\right| .\;\;\;$
\end{proposition}

Our next goal is to estimate the size of $\mathcal{F}(\mathbf{\theta },%
\mathcal{A})$ for a fixed $\mathbf{\theta }\in \Theta _{\varepsilon
,x_{max},n,c}$ and $\mathcal{A\in }\left\{ 0,1\right\} ^{n}$. Abundant use
will be made of the quantities $\tau =\tau (\mathbf{\theta }%
,x_{max})=1-\sum_{0\leq p\leq x\leq x_{max}}\theta _{x,p}$ and $\sigma
=\sigma (\mathbf{\theta },x_{max})=\lambda -\sum_{0\leq p\leq x\leq
x_{max}}x\theta _{x,p}$. $\tau $ is, of course, nonnegative by definition;
for any $F\in \mathcal{F}(\mathbf{\theta },\mathcal{A})$, $\tau $ represents
the proportion of variables having more than $x_{max}$ occurrences in $F$.
Also, $\sigma $ is nonnegative, since for $F\in \mathcal{F}(\mathbf{\theta },%
\mathcal{A})$, it represents the proportion of literals in $F$ (among the
total $\lambda n$) whose underlying variables have more than $x_{max}$
occurrences.

Given a formula $F$ and a truth assignment $\mathcal{A}$, we say that a
clause of $F$ is of type $(\mathcal{A},j)$ ($0\leq j\leq 3$) iff it has $j$
nonzero literals under $\mathcal{A}$. To say that $\mathcal{A}$ satisfies $F$
means that $F$ has no clauses of type $(\mathcal{A},0).$

Now suppose $\mathcal{A}$ is a PPS of $F\in \mathcal{F}(\mathbf{\theta },%
\mathcal{A}),$ let $v$ be one of the variables having $x$ total, $p$
positive occurrences in $F,$ and $q$ its number of occurrences in type-$(%
\mathcal{A},1)$ clauses as the unique satisfying literal. If $v$ has value $%
1 $ under $\mathcal{A},$ then $q=p-j$ for some $j$ with $0\leq j\leq p-1;$
excluding $j=p$ expresses exactly that $\mathcal{A}$ is a PPS. If $v$ has
value $0$ under $\mathcal{A},$ then $q=j-p$ for some $j$ with $p\leq j\leq
x.\;\func{Si}$nce the two cases cover exactly once each possible $j$ between
$0$ and $x,$ they can be conveniently coalesced by saying that for any
variable there is a unique $j$ with $0\leq j\leq x,$ such that $\left|
p-j\right| $ of its occurrences are in type-$(\mathcal{A},1)$ clauses, the
value of the variable under $\mathcal{A}$ being then automatically
determined by the sign of $p-j$. It is $1/2\left( 1+\left( p-j\right)
/\left| p-j\right| \right) $ if $j\neq p$ and by convention $0$ if $j=p$. We
call such a variable a variable of type $(\mathcal{A},x,p,j)$, and thus to
say that $\mathcal{A}$ is a PPS of $F\in \mathcal{F}(\mathbf{\theta },%
\mathcal{A})$ means exactly that every variable is of type $(\mathcal{A}%
,x,p,j)$ for some $x,$ $p$ and $j$ with $0\leq p\leq x$ and $0\leq j\leq x.$
In our enumerations, however, we will only impose this condition for $x\leq
x_{max}.\;$The variables with more than $x_{max}$ occurrences, or \textit{%
heavy} variables, will be considered unconstrained, and we will broadly
overestimate the number of corresponding choices. If our expectation
calculated by excess tends to $0,$ so does the true expectation.

Recall that we use the notation $I_{n}=\left\{ 0,1/n,2/n,...,1-1/n,1\right\}
.$ Given the vector $\mathbf{\theta \in }\Theta _{\varepsilon ,x_{max},n,c},$
the assignment $\mathcal{A},$ and rationals $\gamma _{1},\gamma _{2},\gamma
_{3}\in I_{cn}$ and $\mu _{x,p,j}\in I_{\theta _{x,p}n}\;\left( 0\leq
p,j\leq x\leq x_{max}\right) ,\ $we proceed to count the formulae in $%
\mathcal{F}(\mathbf{\theta },\mathcal{A})$

\begin{itemize}
\item  consisting of $\gamma _{i}cn$ clauses of type $i,\;i=1,2,3,$ and

\item  such that the number of variables of type $(\mathcal{A},x,p,j)$ is $%
\mu _{x,p,j}\theta _{x,p}n$ for $0\leq p,j\leq x\leq x_{max}.$
\end{itemize}

We assume, of course, $\gamma _{1}+\gamma _{2}+\gamma _{3}=1$ and $%
\sum_{j=0}^{x}\mu _{x,p,j}=1$ for $0\leq p,j\leq x\leq x_{max}.$ Let $%
Z\left( \mathbf{\theta },\mathbf{\gamma },\mathbf{\mu },n,c\right) $ be the
number of such formulae.

The empty formula template $\Lambda _{c,n}$ contains $\lambda n$ cells, with
$\lambda =3c.\;$We first choose those which will correspond to each type of
clause, and within each group, those to be filled with literals of value $%
1.\;$This can be done in $A_{n}\left( \mathbf{\gamma },c\right) $ ways,
where
\[
A_{n}\left( \mathbf{\gamma },c\right) =\frac{\left( cn\right) !}{\left(
\gamma _{1}cn\right) !\left( \gamma _{2}cn\right) !\left( \gamma
_{3}cn\right) !}3^{\left( \gamma _{1}+\gamma _{2}\right) cn}.
\]
Second, among the $n$ variables we choose, for$\;0\leq p\leq x\leq x_{max},$
the $\theta _{x,p}n$ which will have $x$ total, $p$ positive occurrences,
and among these the $\mu _{x,p,j}\theta _{x,p}n$ which will be of type $(%
\mathcal{A},x,p,j).\;$Recall that given $\mu _{x,p,j}$ the values under $%
\mathcal{A}$ of the $\mu _{x,p,j}\theta _{x,p}n$ corresponding variables are
automatically determined. We complete the specification of $\mathcal{A}$ by
choosing the values of the remaining $\tau n$ heavy variables (recall $\tau
=1-\sum_{0\leq p\leq x\leq x_{max}}\theta _{x,p}$). The number of
possibilities is:
\[
B_{n}\left( \mathbf{\theta },\mathbf{\mu }\right) =2^{\tau }\frac{n!}{\left(
\tau n\right) !\prod_{0\leq p\leq x\leq x_{max}}\left( \theta _{x,p}n\right)
!}\prod_{0\leq p\leq x\leq x_{max}}\frac{\left( \theta _{x,p}n\right) !}{%
\prod_{j=0}^{x}\left( \mu _{x,p,j}\theta _{x,p}n\right) !}.
\]
Finally, we effectively fill the cells with the variables of different
types.\ Let $M_{n}\left( \mathbf{\theta },\mathbf{\gamma },\mathbf{\mu }%
\right) $ be the number of ways to do this and obtain a formula in $\mathcal{%
F}(\mathbf{\theta },\mathcal{A})$ meeting our requirements. We start with
the heavy variables, which must have $\sigma n$ occurrences (recall $\sigma
=1-\sum_{0\leq p\leq x\leq x_{max}}x\theta _{x,p}$). We assign their
occurrences to cells, which automatically determines the sign of each
occurrence, having already completely specified $\mathcal{A}$ on the one
hand, and the contents, $0$ or $1,$ of each cell, on the other. We bound the
ways to assign all the occurrences of heavy variables to cells by the
quantity
\[
\eta \left( \mathbf{\theta },n,c\right) ={{{{{{{\binom{\lambda n }{\sigma n}}%
}}}}}}\left( \tau n\right) ^{\sigma n}.
\]
The $\gamma _{1}cn$ clauses of type $(\mathcal{A},1)$ contain $\gamma
_{1}\lambda n$ cells, $\gamma _{1}cn$ of which are already reserved for
nonzero literals.\ Among these, some already contain occurrences of heavy
variables. Let their number be $\widehat{\sigma }_{1}n;$ this is not an
independent parameter since
\begin{equation}
\widehat{\sigma }_{1}=\gamma _{1}c-\sum_{0\leq p\leq x\leq x_{max}}\theta
_{x,p}\sum_{0\leq j\leq x}\left| p-j\right| \mu _{x,p,j}.
\label{elimsigma1h}
\end{equation}
There remain $\gamma _{1}c-\widehat{\sigma }_{1}$ cells to be filled in this
group. These are filled with the $p-j$ unnegated occurrences of variables of
type $(\mathcal{A},x,p,j)$ with $0\leq j\leq p-1$ and the $j-p$ negated
occurrences of variables of type $(\mathcal{A},x,p,j)$ with $p\leq j\leq x.$
Thus the number of ways to fill the $\gamma _{1}c-\widehat{\sigma }_{1}$
cells is :
\[
\mathcal{M}_{1}=\frac{\left[ \left( \gamma _{1}c-\widehat{\sigma }%
_{1}\right) n\right] !}{\prod_{0\leq p\leq x\leq x_{max}}\left[
\prod_{j=0}^{p-1}\left( p-j\right) !^{\mu _{x,p,j}\theta
_{x,p}n}\prod_{j=p}^{x}\left( j-p\right) !^{\mu _{x,p,j}\theta
_{x,p}n}\right] }.
\]
Next, we fill the cells already reserved for nonzero literals, which do not
pertain to clauses of type $(\mathcal{A},1).$ It will be convenient to
introduce the normalized nonzero \textit{spread} of $F$ under $\mathcal{A},$
namely
\begin{equation}
\psi =1/3\;\left( \gamma _{1}+2\gamma _{2}+3\gamma _{3}\right) .
\label{defphi}
\end{equation}
Among the $\lambda \psi n$ cells in total which are to receive nonzero
literals, let $\sigma _{1}n$ ones contain occurrences of heavy variables.\ $%
\sigma _{1},$ like $\widehat{\sigma }_{1},$ is a known quantity:
\begin{equation}
\sigma _{1}=\lambda \psi -\sum_{0\leq p\leq x\leq x_{max}}\theta
_{x,p}\left[ p\sum_{0\leq j\leq p-1}\mu _{x,p,j}+\left( x-p\right)
\sum_{p\leq j\leq x}\mu _{x,p,j}\right] ,  \label{elimsigma1}
\end{equation}
For the $\lambda \psi n-\left( \sigma _{1}-\widehat{\sigma }_{1}\right)
n-\gamma _{1}cn$ remaining cells in this group, we have available, for each
variable of type $(\mathcal{A},x,p,j)$ with $0\leq j\leq p-1$, the $p$
unnegated occurrences less $p-j$ already placed; and if the type is $(%
\mathcal{A},x,p,j)$ with $p\leq j\leq x,$ the $x-p$ negated occurrences less
$j-p$ already placed. Thus, the number of ways to do the assignement is
\[
\mathcal{M}_{2}=\frac{\left[ \left( \lambda \psi -\gamma _{1}c-\sigma _{1}+%
\widehat{\sigma }_{1}\right) n\right] !}{\prod_{0\leq p\leq x\leq
x_{max}}\left[ \prod_{j=0}^{p-1}j!^{\mu _{x,p,j}\theta
_{x,p}n}\prod_{j=p}^{x}\left( x-j\right) !^{\mu _{x,p,j}\theta
_{x,p}n}\right] }.
\]
Lastly, we deal with the $\lambda \left( 1-\psi \right) n$ cells reserved
for null literals, of which $\left( \sigma -\sigma _{1}\right) n$ are
already filled. For the remaining ones, we have $x-p$ occurrences of
variables of type $(\mathcal{A},x,p,j)$ with $0\leq j\leq p-1,$ and $p$
occurrences if $p\leq j\leq x.\;$So, we can fill them in $\mathcal{M}_{3}$
ways, where
\[
\mathcal{M}_{3}=\frac{\left\{ \left[ \lambda \left( 1-\psi \right) -\sigma
+\sigma _{1}\right] n\right\} !}{\prod_{0\leq p\leq x\leq x_{max}}\left[
\left( x-p\right) !^{\sum_{j=0}^{p-1}\mu _{x,p,j}\theta
_{x,p}n}\;p!^{\sum_{j=p}^{x}\mu _{x,p,j}\theta _{x,p}n}\right] }
\]
To sum up, $M_{n}\left( \mathbf{\theta },\mathbf{\gamma },\mathbf{\mu }%
\right) \leq \mathcal{M}_{1}\mathcal{M}_{2}\mathcal{M}_{3}\;\eta \left(
\mathbf{\theta },n,c\right) ,$ so that
\begin{equation}
Z\left( \mathbf{\theta },\mathbf{\gamma },\mathbf{\mu },n,c\right) \leq
A_{n}\left( \mathbf{\gamma },c\right) B_{n}\left( \mathbf{\theta },\mathbf{%
\mu }\right) \;\mathcal{M}_{1}\mathcal{M}_{2}\mathcal{M}_{3}\;\eta \left(
\mathbf{\theta },n,c\right) .  \label{majZ}
\end{equation}

\subsection{The expectation.}

It follows from (\ref{majZ}), the preceding discussion, and the definition
of $\mathcal{F}(\mathbf{\theta },\mathcal{A}),$ that, setting $%
J_{n}=\bigcup_{1\leq k\leq n}I_{k},$ we have
\begin{equation}
\sum_{\mathcal{A}\in \left\{ 0,1\right\} ^{n}}\left| \mathcal{F}(\mathbf{%
\theta },\mathcal{A})\right| \leq \eta \left( \mathbf{\theta },n,c\right)
\sum_{\mathbf{\gamma }\in I_{cn}}A_{n}\left( \mathbf{\gamma },c\right) \sum_{%
\mathbf{\mu }\in J_{n}}B_{n}\left( \mathbf{\theta },\mathbf{\mu }\right) \;%
\mathcal{M}_{1}\mathcal{M}_{2}\mathcal{M}_{3},  \label{Ftha_bound}
\end{equation}
where the summation is under the constraints
\begin{equation}
\gamma _{1}+\gamma _{2}+\gamma _{3}=1  \label{sumbeta}
\end{equation}
and
\begin{equation}
\sum_{j=0}^{x}\mu _{x,p,j}=1,\;0\leq p\leq x\leq x_{max},  \label{summu}
\end{equation}
and where $\sigma _{1}$ and $\widehat{\sigma }_{1}$ are expressed in $%
\mathcal{M}_{1},\mathcal{M}_{2}$ and $\mathcal{M}_{3}$ as functions of
\textbf{$\gamma $} and $\mathbf{\mu }$, see (\ref{elimsigma1}) and (\ref
{elimsigma1h}).

We now introduce a modified form of (\ref{elimsigma1}) which will be
convenient later. We set
\begin{equation}
\alpha _{x,p}=\sum_{0\leq j\leq p-1}\mu _{x,p,j},  \label{alphadef}
\end{equation}
the proportion of variables with $x$ total, $p$ positive occurrences having
the value $1$ under $\mathcal{A}.$ Taking account of (\ref{summu}), (\ref
{elimsigma1}) can be written using only the $\alpha _{x,p}$'s:
\begin{equation}
\lambda \psi -\sigma _{1}=K\left( \mathbf{\theta }\right) \mathbf{-}%
\sum_{0\leq p\leq x\leq x_{max}}H_{x,p}\left( \mathbf{\theta }\right) \alpha
_{x,p},\;\;\;\;\mathrm{where}  \label{sigma1_mufree}
\end{equation}
\[
K\left( \mathbf{\theta }\right) =\sum_{0\leq p\leq x\leq x_{max}}\left(
x-p\right) \theta _{x,p}\;\;\mathrm{and,\;for\;}0\leq p\leq x\leq
x_{max}:\;\;H_{x,p}\left( \mathbf{\theta }\right) =\left( x-2p\right) \theta
_{x,p}.
\]
From Lemma \ref{thetasize}, Proposition \ref{Efromfixtheta}, and (\ref
{Ftha_bound}), we get, for any fixed $\mathbf{\theta }\in \Theta
_{\varepsilon ,x_{max},n,c}:$%
\begin{equation}
\mathbf{E}\left( X_{n,\varepsilon ,x_{max},c}\right) \leq \frac{%
2^{n\sum_{x>2p}\tilde{\kappa}_{x,p}}}{\left( 2n\right) ^{\lambda n}}\left(
2\varepsilon n\right) ^{D}\eta \left( \mathbf{\theta },n,c\right) \sum_{%
\mathbf{\gamma }\in I_{cn}}A_{n}\left( \mathbf{\gamma },c\right) \sum_{%
\mathbf{\mu }\in J_{n}}B_{n}\left( \mathbf{\theta },\mathbf{\mu }\right) \;%
\mathcal{M}_{1}\mathcal{M}_{2}\mathcal{M}_{3},  \label{firstEbound}
\end{equation}
subject again to (\ref{sumbeta}) and (\ref{summu}).

\section{Asymptotics.}

\subsection{Bound for the exponential order.}

Still for a fixed $\mathbf{\theta },$ we now bound the general term of (\ref
{Ftha_bound}), using a standard inequality for multinomial coefficients:
\[
\left(
\begin{array}{c}
r \\
r_{1}\;r_{2}\;...\;r_{s}
\end{array}
\right) \leq \frac{r^{r}}{r_{1}^{r_{1}}r_{2}^{r_{2}}...r_{s}^{r_{s}}},
\]
which gives first, taking account of (\ref{sumbeta}):
\[
A_{n}(\mathbf{\gamma },c)^{1/n}\leq \frac{3^{c}}{\left[ \gamma _{1}^{\gamma
_{1}}\gamma _{2}^{\gamma _{2}}\left( 3\gamma _{3}\right) ^{\gamma
_{3}}\right] ^{c}};
\]
further,
\[
B_{n}(\mathbf{\theta },\mathbf{\mu })^{1/n}\leq 2^{\tau }\frac{1}{\tau
^{\tau }\prod\limits_{0\leq p\leq x\leq x_{max}}\theta _{x,p}^{\theta
_{x,p}}\prod\limits_{0\leq p\leq x\leq x_{max}}\left(
\prod\limits_{j=0}^{x}\mu _{x,p,j}^{\mu _{x,p,j}}\right) ^{\theta _{x,p}}};
\]
\[
\mathcal{M}_{1}{}^{1/n}\leq \frac{\left( \gamma _{1}c-\widehat{\sigma }%
_{1}\right) ^{\gamma _{1}c-\widehat{\sigma }_{1}}\left( \frac{n}{e}\right)
^{\gamma _{1}c-\widehat{\sigma }_{1}}}{\prod\limits_{0\leq p\leq x\leq
x_{max}}\left[ \prod\limits_{j=0}^{p-1}\left( p-j\right) !^{\mu
_{x,p,j}\theta _{x,p}}\prod\limits_{j=p}^{x}\left( j-p\right) !^{\mu
_{x,p,j}\theta _{x,p}}\right] };
\]
next, bearing in mind that, by (\ref{sigma1_mufree})\textit{, }$\lambda \psi
-\sigma _{1}$ does not depend on $n$:
\[
\mathcal{M}_{2}{}^{1/n}\leq \frac{\left[ \left( \lambda \psi -\gamma
_{1}c-\sigma _{1}+\widehat{\sigma }_{1}\right) \frac{n}{e}\right] ^{\lambda
\psi -\gamma _{1}c-\sigma _{1}+\widehat{\sigma }_{1}}}{\prod\limits_{0\leq
p\leq x\leq x_{max}}\left[ \prod\limits_{j=0}^{p-1}j!^{\mu _{x,p,j}\theta
_{x,p}}\prod\limits_{j=p}^{x}\left( x-j\right) !^{\mu _{x,p,j}\theta
_{x,p}}\right] };
\]
and, using (\ref{alphadef}):
\[
\mathcal{M}_{3}^{1/n}\leq \frac{\left\{ \left[ \lambda \left( 1-\psi \right)
+\sigma _{1}-\sigma \right] \frac{n}{e}\right\} ^{\lambda \left( 1-\psi
\right) +\sigma _{1}-\sigma }}{\prod\limits_{0\leq p\leq x\leq
x_{max}}\left[ \left( x-p\right) !^{\alpha _{x,p}}p!^{1-\alpha
_{x,p}}\right] ^{\theta _{x,p}}}.
\]
So, writing $\left( p-j\right) !j!=p!/{{{{{{{\binom{p}{j}}}}}}}}$ and $%
\left( j-p\right) !\left( x-j\right) !=\left( x-p\right) !/{{{{{{{\binom{x-p%
}{j-p}}}}}}}}:$%
\begin{eqnarray*}
\mathcal{M}_{1}{}^{1/n}\mathcal{M}_{2}{}^{1/n} &\leq &\left( \gamma _{1}c-%
\widehat{\sigma }_{1}\right) ^{\gamma _{1}c-\widehat{\sigma }_{1}}\left[
\left( \lambda \psi -\gamma _{1}c-\sigma _{1}+\widehat{\sigma }_{1}\right)
\frac{n}{e}\right] ^{\lambda \psi -\gamma _{1}c-\sigma _{1}+\widehat{\sigma }%
_{1}}\times  \\
&&\left( \frac{n}{e}\right) ^{\gamma _{1}c-\widehat{\sigma }%
_{1}}\prod\limits_{0\leq p\leq x\leq x_{max}}\frac{\left[
\prod\limits_{j=0}^{p-1}{{{{{{{\binom{p}{j}}}}}}}}^{\mu
_{x,p,j}}\prod\limits_{j=p}^{x}{{{{{{{\binom{x-p}{j-p}}}}}}}}^{\mu
_{x,p,j}}\right] ^{\theta _{x,p}}}{\left[ p!^{\alpha _{x,p}}\left(
x-p\right) !^{1-\alpha _{x,p}}\right] ^{\theta _{x,p}}}.
\end{eqnarray*}
Bounding the sum in (\ref{Ftha_bound}) by its maximum term times the number
of terms $\left| I_{cn}\right| \left| J_{n}\right| $ with $\left|
J_{n}\right| \leq n\left( n+3\right) /2$, we get, after some simplification
and whenever $c\leq 5$:

\[
\left[ \sum_{\mathcal{A\in }\left\{ 0,1\right\} ^{n}}\left| \mathcal{F}(%
\mathbf{\theta },\mathcal{A})\right| \right] ^{\frac{1}{n}}\leq \frac{%
3^{c}\left( \lambda n\right) ^{\lambda }}{\sigma ^{\sigma }}\left[ \frac{c}{%
\left( \lambda -\sigma \right) e}\right] ^{\lambda -\sigma }\frac{\tau
^{\sigma }.\left( 2/\tau \right) ^{\tau }}{\prod\limits_{0\leq p\leq x\leq
x_{max}}\left[ p!\left( x-p\right) !\theta _{x,p}\right] ^{\theta _{x,p}}}%
\times
\]
\begin{eqnarray}
&&\;\;\;\;\;\;\;\;\;\;\;\;\;\;\;\;\;\;\;\;\;\;\;\;\;\;\;\;\;\;\;\;\;\;\;\;\;%
\;\;\;\;\left( 6n^{3}\right) ^{\frac{1}{n}}\max_{\gamma _{j}\in I_{cn},%
\mathrm{\ }\mu _{x,p,j}\in J_{n}}\frac{\left( \gamma _{1}-\widehat{\sigma }%
_{1}/c\right) ^{\gamma _{1}c-\widehat{\sigma }_{1}}}{\left[ \gamma
_{1}^{\gamma _{1}}\gamma _{2}^{\gamma _{2}}\left( 3\gamma _{3}\right)
^{\gamma _{3}}\right] ^{c}}\times   \label{thbound1} \\
&&\;\;\;\;\;\;\;\;\;\frac{\left[ 3\left( 1-\psi \right) +\sigma
_{1}/c-\sigma /c\right] ^{\lambda \left( 1-\psi \right) +\sigma _{1}-\sigma
}\left( 3\psi -\gamma _{1}-\sigma _{1}/c+\widehat{\sigma }_{1}/c\right)
^{\lambda \psi -\gamma _{1}c-\sigma _{1}+\widehat{\sigma }_{1}}}{%
\prod\limits_{0\leq p\leq x\leq x_{max}}\left[ \prod\limits_{j=0}^{x}\left(
\frac{\mu _{x,p,j}}{h_{x,p,j}}\right) ^{\mu _{x,p,j}}\right] ^{\theta _{x,p}}%
},  \nonumber
\end{eqnarray}
where the maximum is subject to all the above constraints (\ref{sumbeta})%
\textit{\ }and (\ref{summu})\textit{,} and where
\[
h_{x,p,j}=\left\{
\begin{array}{lll}
{{{\binom{p}{j}}}} & \mathrm{if} & 0\leq j\leq p-1, \\
&  &  \\
{{{\binom{x-p}{j-p}}}} & \mathrm{if} & p\leq j\leq x.
\end{array}
\right.
\]
Since ${{{{{{{\binom{p}{j}}}}}}}}={{{{{{{\binom{p}{p-j}}}}}}}}$, it may be
observed that $h_{x,p,j}$ is the number of ways to select , among the
\textit{literals} with value $1$ under $\mathcal{A}$ associated with a given
variable of type $(\mathcal{A},x,p,j)$ (assumed distinguishable), those (if
any) destined to prevent the flipping of that variable.

Finally, still for a fixed value of $\mathbf{\theta }\in \Theta
_{\varepsilon ,x_{max},n,c}$, we can extend the $\max $ in the above
estimate to arbitrary real values of the $\gamma _{j}$'s and of the $\mu
_{x,p,j}$'s in $\left[ 0,1\right] $, subject to the stated constraints.

\subsection{\textit{A priori} bounds on the main parameters.\label%
{a_pr_bounds}}

We are about to replace our estimate (\ref{thbound1}) by one that is
uniformly valid for all $\mathbf{\theta \in }\Theta _{\varepsilon
,x_{max},n,c},$ and to that end will require that $c$ be bounded from above
and below, and will have to check some inequalities involving $c,\varepsilon
$ and $x_{max}.$ To give our estimate in reasonable generality, we assume $%
0<c_{min}\leq c\leq c_{max}$ with, for the moment, only a mild and fairly
arbitrary constraint on $c_{min}$ and $c_{max},$ say $3\leq c_{min}\leq
c_{max}\leq 5;$ correspondingly, $\lambda $ is restricted to $\left[ \lambda
_{min},\lambda _{max}\right] $ with $9\leq $ $\lambda _{min}\leq \lambda
_{max}\leq 15.$ Later, we will be more specific and impose $%
c_{min}=c=c_{max}=4.506.$

For such an interval $[c_{min},c_{max}],$ it is easy, by elementary
expectation calculations, to determine intervals $\left[ \gamma
_{1min},\gamma _{1max}\right] ,\left[ \gamma _{2min},\gamma _{2max}\right]
,\left[ \gamma _{3min},\gamma _{3max}\right] ,\left[ \psi _{min},\psi
_{max}\right] ,$ such that for $c\in [c_{min},c_{max}],$ the probability
that a formula in $\Omega (n,c)$ has a solution with \textit{at least one}
of $\gamma _{1},\gamma _{2},\gamma _{3,}\psi $ falling \textit{outside} the
corresponding range is \textit{always} exponentially small. For example, for
$[c_{min},c_{max}]\subset \left[ 3,5\right] $ we can take these intervals to
be $\left[ 0.21,0.65\right] ,\left[ 0.21,0.65\right] ,\left[
0.017,0.32\right] ,$ and $\left[ 0.47,0.68\right] ,$ respectively.

This means that in investigating, by more sophisticated means, the
probability that a formula in $\Omega (n,c)$ is satisfiable, we need only
consider solutions, or indeed PPSs, with $\gamma _{1},\gamma _{2},\gamma
_{3,}\;$and $\psi $ in their respective intervals. Thus, we can define the
r.v. $X_{n,\varepsilon ,x_{max},c}$ with these more restricted PPSs, and $%
\mathcal{F}(\mathbf{\theta },\mathcal{A})$ similarly.\ All that we have said
up to now goes over, notably Propositions \ref{domtyp} and \ref
{Efromfixtheta}; and (\ref{thbound1}) holds, with the maximum subject to
these additional restrictions, viz
\begin{equation}
\gamma _{j}\in \left[ \gamma _{j\;min},\gamma _{j\;max}\right]
,\;j=1,2,3;\;\;\;\;\;\;\psi \in \left[ \psi _{min},\psi _{max}\right] .
\label{GamPsiConstr}
\end{equation}

Henceforth we assume these additional constraints throughout; we also fix $%
\varepsilon =10^{-15}$ and $x_{max}=56.$

\subsection{The $\mathbf{\theta }$-free estimate.\label{tfe}}

Deriving from (\ref{thbound1}), at controllably small cost, an estimate
where the fixed but unknown $\theta _{x,p}$'s are replaced by the known $%
\widetilde{\kappa }_{x,p}$'s, is a matter of easy but tedious calculations
which we will only sketch.\ Anyway, one could get by with coarser bounds
than we give by simply choosing a smaller $\varepsilon $ and larger $x_{max}.
$ Note that we have relied on $x_{max}$ being even to simplify some of the
calculations slightly.

One somewhat delicate point is how to deal with the numerous quantities of
the form $x^{x}$ where $x$ is unknown but `near' some known $y.$ We use the
following very elementary lemma, not sharp but sufficient, so not worth
improving.

\begin{lemma}
\label{explemma} Let $L$ be the function $\eta \mapsto \left( 2\eta \right)
^{-2\eta }$ on $\Bbb{R}_{+}$. Then whenever $x,y,\eta $ are positive reals
with $\eta \leq 0.05$, $\left| x-y\right| \leq \eta $, and $x\leq 30$, we
have $L\left( \eta \right) ^{-1}\leq y^{y}/x^{x}\leq L\left( \eta \right) $.
\end{lemma}

\begin{proof}
(outline)\ For $x\leq 30$, we study the function $f_{x}\left( h\right)
=\left( x+h\right) ^{x+h}$ on the interval $I_{\eta }\left( x\right) =\left[
\max \left( -x,-\eta \right) ,\eta \right] $, showing that whenever $\left|
h\right| \leq \eta ,\;f_{x}\left( h\right) /x^{x}$ falls between $L\left(
\eta \right) ^{-1}$ and $L\left( \eta \right) $. This is done by elementary
monotony considerations, distinguishing the two cases $\left| h\right| >x$
and $\left| h\right| \leq x$, the second being split into two subcases where
the double inequality $1/e-\eta \leq x\leq 1/e+\eta $ either holds or not.
\end{proof}

\subsubsection{Eliminating the $\sigma $'s and $\tau $, and withdrawing $%
\mathbf{\theta }$ from $\psi $ and $\mathbf{\gamma }$.}

From $\tau \leq 1-\sum_{x,p}\widetilde{\kappa
}_{x,p}+\sum_{x,p}\left|
\theta _{x,p}-\widetilde{\kappa }_{x,p}\right| $, we get, in terms of $D=$ $%
D\left( x_{max}\right) $,
\[
\tau \left( \mathbf{\theta ,}x_{max},\lambda \right) \leq R_{1}\left(
\varepsilon ,x_{max}\right)
\]
uniformly for all $\mathbf{\theta }\in \Theta _{\varepsilon ,x_{max},n,c}$
and all relevant $\lambda ,$ where
\[
R_{1}\left( \varepsilon ,x_{max}\right) =\frac{\lambda _{max}^{^{x_{max}+1}}%
}{\left( x_{max}+1\right) !}+\varepsilon D\left( x_{max}\right) .
\]
Similarly, from $\sigma \leq \left( \lambda -\sum_{x,p}x\widetilde{\kappa }%
_{x,p}\right) +\sum_{x,p}x\left| \theta _{x,p}-\widetilde{\kappa }%
_{x,p}\right| $ we obtain, in terms of $P_{2}\left( \xi \right) =\xi \left(
\xi +1\right) \left( \xi +2\right) /3$ that $\sigma \left( \mathbf{\theta ,}%
x_{max},\lambda \right) \leq R_{2}\left( \varepsilon ,x_{max}\right) ,$
again uniformly for all $\mathbf{\theta }\in \Theta _{\varepsilon
,x_{max},n,c}$ and relevant $\lambda ,$ where
\[
R_{2}\left( \varepsilon ,x_{max}\right) =\frac{\lambda _{max}^{^{x_{max}+1}}%
}{x_{max}!}+\varepsilon P_{2}\left( x_{max}\right) .
\]
As for $\sigma _{1}$ and $\widehat{\sigma }_{1},$ we simply use $\widehat{%
\sigma }_{1}\leq \sigma _{1}\leq \sigma .$

Turning to $\psi $ and $\mathbf{\gamma ,}$ from (\ref{elimsigma1h}) and (\ref
{elimsigma1}) there are natural candidates for the $\theta $-free versions
of $\gamma _{1}$ and $\psi ,$ namely
\begin{equation}
\beta _{1}=\frac{1}{c}\sum_{0\leq p\leq x\leq x_{max}}\widetilde{\kappa }%
_{x,p}\left[ \sum_{0\leq j\leq p-1}\left( p-j\right) \mu
_{x,p,j}+\sum_{p\leq j\leq x}\left( j-p\right) \mu _{x,p,j}\right] ,
\label{defgamma1}
\end{equation}
\begin{equation}
\phi =\frac{1}{\lambda }\sum_{x,p}\widetilde{\kappa }_{x,p}\left[
p\sum_{0\leq j\leq p-1}\mu _{x,p,j}+\left( x-p\right) \sum_{p\leq j\leq
x}\mu _{x,p,j}\right] ,  \label{defpsi}
\end{equation}
or, equivalently so long as (\ref{summu}) holds, cf. (\ref{sigma1_mufree}):
\begin{equation}
\phi =\frac{1}{\lambda }\left[ \widetilde{K}-\sum_{0\leq 2p\leq x\leq
x_{max}}\widetilde{H}_{x,p}\alpha _{x,p}\right]  \label{defphimod}
\end{equation}
with
\[
\widetilde{K}=\sum_{0\leq 2p\leq x\leq x_{max}}\left( x-p\right) \widetilde{%
\kappa }_{x,p}\;\;\;\;\;\;\mathrm{and}\;\;\;\;\;\;\widetilde{H}_{x,p}=\left(
x-2p\right) \widetilde{\kappa }_{x,p},
\]
which is the definition of $\phi $ we adopt, since it will be helpful
subsequently.

Observing that from (\ref{sumbeta}) and (\ref{defphi}), we have $\gamma
_{2}=3\left( 1-\phi \right) -2\gamma _{1}$ and $\gamma _{3}=\gamma
_{1}-2+3\phi ,$ we \textit{define} the $\mathbf{\theta }$-free versions of $%
\gamma _{2}$ and $\gamma _{3}$ as respectively
\begin{equation}
\beta _{2}=3\left( 1-\psi \right) -2\beta _{1}\;\;\;\;\mathrm{and}%
\;\;\;\;\beta _{3}=\beta _{1}-2+3\psi .  \label{defb23}
\end{equation}

Using the above bounds on $\sigma $, it is easy to estimate $\left| \phi
-\psi \right| ,\left| \beta _{j}-\gamma _{j}\right| ,$ and the worst-case
error incurred in replacing $\phi $ by $\psi $ and the $\beta _{j}$'s by the
$\gamma _{j}$'s in (\ref{thbound1}). Setting $P_{3}\left( \xi \right) =\xi
\left( \xi +2\right) \left( 2\xi +3\right) /8$ and
\[
R_{3}\left( \varepsilon ,x_{max}\right) =\frac{\lambda _{max}^{^{x_{max}}}}{%
x_{max}!}+\frac{\varepsilon }{\lambda _{min}}\left[ P_{2}\left(
x_{max}\right) +P_{3}\left( x_{max}\right) \right] ,
\]
we find
\begin{equation}
\left.
\begin{array}{ll}
\left| \phi -\psi \right| \leq R_{3}\left( \varepsilon ,x_{max}\right) , &
\left| \beta _{1}-\gamma _{1}\right| \leq 3R_{3}\left( \varepsilon
,x_{max}\right) , \\
\left| \beta _{2}-\gamma _{2}\right| \leq 9R_{3}\left( \varepsilon
,x_{max}\right) , & \left| \beta _{3}-\gamma _{3}\right| \leq 6R_{3}\left(
\varepsilon ,x_{max}\right) ,
\end{array}
\right\}  \label{deltaphib123}
\end{equation}
and therefore the constraints (\ref{GamPsiConstr}) imply the following ones
on $\mathbf{\mu }:$%
\begin{equation}
\beta _{j}\in \left[ \beta _{j\;min},\beta _{j\;max}\right]
,\;j=1,2,3;\;\;\;\;\;\;\phi \in \left[ \phi _{min},\phi _{max}\right]
\label{BetaPhiConstr}
\end{equation}
where $\beta _{1min}=\gamma _{1min}-3R_{3}\left( \varepsilon ,x_{max}\right)
,\beta _{1max}=\gamma _{1max}+3R_{3}\left( \varepsilon ,x_{max}\right) ,$
and so on.

Since $R_{3}\left( \varepsilon ,x_{max}\right) <1.035\;10^{-9},$ from (\ref
{deltaphib123}) and (\ref{BetaPhiConstr}) we see that all of $\beta
_{1},\beta _{2},\beta _{3},\phi ,3\phi -\beta _{1},$ and $3\left( 1-\phi
\right) $ are \textit{positive,} $\leq 30,$ and at a distance of less than $%
0.05$ from the corresponding $\mathbf{\theta }$-dependent quantities. This
allows us, using also $\sigma \leq R_{2}\left( \varepsilon ,x_{max}\right)
<1.54\;10^{-8},$ repeatedly to apply Lemma \ref{explemma} and get:
\begin{eqnarray*}
\left[ \sum_{\mathcal{A\in }\left\{ 0,1\right\} ^{n}}\left| \mathcal{F}(%
\mathbf{\theta },\mathcal{A})\right| \right] ^{\frac{1}{n}} &\leq
&G_{1}\left( \varepsilon ,x_{max}\right) 3^{c}\left( \lambda n\right)
^{\lambda }\left( \frac{1}{3e}\right) ^{\lambda }\times \\
&&\left( 6n^{3}\right) ^{\frac{1}{n}}\max_{\mu _{x,p,j}\in \left[ 0,1\right]
}\frac{\left\{ \left( 3\phi -\beta _{1}\right) ^{3\phi -\beta _{1}}\left[
3\left( 1-\phi \right) \right] ^{3\left( 1-\phi \right) }\beta _{2}^{-\beta
_{2}}\left( 3\beta _{3}\right) ^{-\beta _{3}}\right\} ^{c}}{%
\prod\limits_{0\leq p\leq x\leq x_{max}}\left[ p!\left( x-p\right) !\theta
_{x,p}\prod\limits_{j=0}^{x}\left( \frac{\mu _{x,p,j}}{h_{x,p,j}}\right)
^{\mu _{x,p,j}}\right] ^{\theta _{x,p}}},
\end{eqnarray*}
where
\begin{eqnarray*}
G_{1}\left( \varepsilon ,x_{max}\right) &=&2^{R_{1}\left( \varepsilon
,x_{max}\right) }L\left( R_{1}\left( \varepsilon ,x_{max}\right) \right)
L\left( R_{2}\left( \varepsilon ,x_{max}\right) \right) \left( \frac{18\;e}{%
\lambda _{min}}\right) ^{R_{2}\left( \varepsilon ,x_{max}\right) }\times \\
&&\left[ L\left( \frac{R_{2}\left( \varepsilon ,x_{max}\right) }{6}\right)
\right] ^{6}\left[ L\left( \frac{3\varepsilon }{\lambda _{min}}P_{3}\left(
x_{max}\right) \right) L\left( \frac{6\varepsilon }{\lambda _{min}}%
P_{3}\left( x_{max}\right) \right) \right] ^{c_{max}}\times \\
&&\left[ \left( L\left( 3R_{3}\left( \varepsilon ,x_{max}\right) \right)
\right) ^{2}L\left( 9R_{3}\left( \varepsilon ,x_{max}\right) \right) L\left(
6R_{3}\left( \varepsilon ,x_{max}\right) \right) 3^{R_{3}\left( \varepsilon
,x_{max}\right) }\right] ^{c_{max}},
\end{eqnarray*}
and the $\max $ is subject to (\ref{summu}) and (\ref{BetaPhiConstr}).

\subsubsection{Removing $\mathbf{\theta }$ from the powers-and-factorials
product.}

Since the only real difficulty is, possibly, getting started on the right
path, we only indicate how we break down the error incurred from replacing
the $\theta _{x,p}$'s by the $\kappa _{x,p}$'s, into three factors $A,B,C,$
to be estimated separately. We have
\[
\frac{1}{\prod\limits_{0\leq p\leq x\leq x_{max}}\left[ p!\left( x-p\right)
!\theta _{x,p}\prod\limits_{j=0}^{x}\left( \frac{\mu _{x,p,j}}{h_{x,p,j}}%
\right) ^{\mu _{x,p,j}}\right] ^{\theta _{x,p}}}\leq \frac{ABC}{%
\prod\limits_{0\leq 2p\leq x\leq x_{max}}\left[ p!\left( x-p\right) !%
\widetilde{\kappa }_{x,p}\prod\limits_{j=0}^{x}\left( \frac{\mu _{x,p,j}}{%
h_{x,p,j}}\right) ^{\mu _{x,p,j}}\right] ^{\widetilde{\kappa }_{x,p}}}
\]
(note the change of domain for the index $p$), with
\begin{eqnarray*}
A &=&\prod\limits_{1\leq p\leq x\leq x_{max},\mathrm{\ }2p>x}\left[ p!\left(
x-p\right) !\theta _{x,p}\prod\limits_{j=0}^{x}\left( \frac{\mu _{x,p,j}}{%
h_{x,p,j}}\right) ^{\mu _{x,p,j}}\right] ^{-\theta _{x,p}}, \\
B &=&\prod_{0\leq 2p\leq x\leq x_{max}}\left( \frac{\widetilde{\kappa }_{x,p}%
}{\theta _{x,p}}\right) ^{\theta _{x,p}}, \\
C &=&\prod_{0\leq 2p\leq x\leq x_{max}}\left[ p!\left( x-p\right) !%
\widetilde{\kappa }_{x,p}\prod\limits_{j=0}^{x}\left( \frac{\mu _{x,p,j}}{%
h_{x,p,j}}\right) ^{\mu _{x,p,j}}\right] ^{\widetilde{\kappa }_{x,p}-\theta
_{x,p}}.
\end{eqnarray*}
We find $A\leq G_{A}\left( \varepsilon ,x_{max}\right) ,\;B\leq G_{B}\left(
\varepsilon ,x_{max}\right) ,\;C\leq G_{C}\left( \varepsilon
,x_{max},\lambda \right) ,$ with
\[
G_{A}\left( \varepsilon ,x_{max}\right) =12^{\varepsilon }2^{\frac{x_{max}}{8%
}\left( x_{max}^{2}+2x_{max}-1\right) \varepsilon }L\left( \varepsilon
\right) ^{\frac{x_{max}}{4}\left( x_{max}+3\right) },
\]
\[
G_{B}\left( \varepsilon ,x_{max}\right) =\prod_{x=0}^{x_{max}}\left(
1+\varepsilon \right) ^{\left\lfloor \frac{x}{2}\right\rfloor }\leq \left(
1+\varepsilon \right) ^{\frac{x_{max}}{4}\left( x_{max}+1\right) },
\]
and, using the fact that
\begin{equation}
\mathrm{for\;all\;}\lambda \in \left[ \lambda _{min},\lambda _{max}\right]
,\ \ \ \ x_{max}\geq \frac{2\lambda -\log 2}{\log \lambda -\log 2},
\label{xmaxlb1}
\end{equation}
\[
G_{C}\left( \varepsilon ,x_{max},\lambda \right)
=\;\;\;\;\;\;\;\;\;\;\;\;\;\;\;\;\;\;\;\;\;\;\;\;\;\;\;\;\;\;\;\;\;\;\;\;\;%
\;\;\;\;\;\;\;\;\;\;
\]
\[
\left\{ \left( x_{max}+1\right) ^{\frac{x_{max}+2}{2}}2^{\frac{x_{max}}{24}%
\left( x_{max}+1\right) \left( x_{max}-7\right) }\left\{ 2^{x_{max}+4}\left[
e^{-\lambda }\left( \frac{\lambda }{2}\right) ^{x_{max}}\right]
^{x_{max}+8}\right\} ^{\frac{x_{max}+1}{4}}\right\} ^{\varepsilon }.
\]
Observing that since
\begin{equation}
x_{max}>\lambda _{max},  \label{xmaxlb2}
\end{equation}
$e^{-\lambda }\left( \frac{\lambda }{2}\right) ^{x_{max}}$ increases with $%
\lambda $ within our range of interest, and setting $G_{2}\left( \varepsilon
,x_{max}\right) =$\newline
$G_{A}\left( \varepsilon ,x_{max}\right) G_{B}\left( \varepsilon
,x_{max}\right) G_{C}\left( \varepsilon ,x_{max},\lambda _{max}\right) ,$ we
conclude that, for any $c$ in our chosen range and the $\max $ being again
subject to (\ref{summu}) and (\ref{BetaPhiConstr}):
\begin{eqnarray}
\left[ \sum_{\mathcal{A\in }\left\{ 0,1\right\} ^{n}}\left| \mathcal{F}(%
\mathbf{\theta },\mathcal{A})\right| \right] ^{\frac{1}{n}} &\leq
&G_{1}\left( \varepsilon ,x_{max}\right) G_{2}\left( \varepsilon
,x_{max}\right) 3^{c}\left( \lambda n\right) ^{\lambda }\left( \frac{1}{3e}%
\right) ^{\lambda }\left( 6n^{3}\right) ^{\frac{1}{n}}\times  \label{tfb1} \\
&&\max_{\mu _{x,p,j}\in \left[ 0,1\right] }\frac{\left\{ \left( 3\phi -\beta
_{1}\right) ^{3\phi -\beta _{1}}\left[ 3\left( 1-\phi \right) \right]
^{3\left( 1-\phi \right) }\beta _{2}^{-\beta _{2}}\left( 3\beta _{3}\right)
^{-\beta _{3}}\right\} ^{c}}{\prod\limits_{0\leq 2p\leq x\leq x_{max}}\left[
p!\left( x-p\right) !\widetilde{\kappa }_{x,p}\prod\limits_{j=0}^{x}\left(
\frac{\mu _{x,p,j}}{h_{x,p,j}}\right) ^{\mu _{x,p,j}}\right] ^{\widetilde{%
\kappa }_{x,p}}}.  \nonumber
\end{eqnarray}
Note that the $\mu _{x,p,j}$'s with $2p>x$ are now irrelevant, having
vanished from the bound (they do not actually figure in $\beta _{1}$, $\beta
_{2}$, $\beta _{3}$, or $\phi $), and thus the equality constraints under
which we now perform the maximisation are just (\ref{summu}) for $0\leq
2p\leq x\leq x_{max}$.

\section{Maximization.}

By (\ref{BetaPhiConstr}), for $c$ within our range, the max on the r.h.s. of
(\ref{tfb1}) may be restricted to vectors $\mathbf{\mu }\in \mathcal{U},$
where
\[
\mathcal{U}=\beta _{1}^{-1}\left( \left] \beta _{1min},\beta _{1max}\right[
\right) \cap \beta _{2}^{-1}\left( \left] \beta _{2min},\beta _{2max}\right[
\right) \cap \beta _{3}^{-1}\left( \left] \beta _{3min},\beta _{3max}\right[
\right) \cap \phi ^{-1}\left( \left] \phi _{min},\phi _{max}\right[ \right)
,
\]
is an open subset of $\Bbb{R}^{N}$ where $N=N\left( x_{max}\right) =1/24$ $%
\left( x_{max}+2\right) \left( 4x_{max}^{2}+13x_{max}+12\right) $ (recall
that we have dropped the irrelevant variables $\mu _{x,p,j}$ with $p>x/2$).
For the moment, we do not further specify these reals; we do so later when
restricting $c$ to a single value. (We do already assume $3\phi _{min}>\beta
_{1max}$ though). For now, (\ref{tfb1}) leads to the following problem of
constrained maximization:
\begin{eqnarray}
&&\max_{\mathbf{\mu }\in \Bbb{R}_{+}^{N}\cap \mathcal{U}}\sum_{0\leq 2p\leq
x\leq x_{max}}\tilde{\kappa}_{x,p}\sum_{j=0}^{x}\mu _{x,p,j}\log \left(
\frac{h_{x,p,j}}{\mu _{x,p,j}}\right)  \label{pbmax1} \\
&&+c\left\{ \left( 3\phi -\beta _{1}\right) \log \left( 3\phi -\beta
_{1}\right) +\left[ 3\left( 1-\phi \right) \right] \log \left[ 3\left(
1-\phi \right) \right] -\beta _{2}\log \beta _{2}-\beta _{3}\log \left(
3\beta _{3}\right) \right\}  \nonumber
\end{eqnarray}
subject to the constraints \textit{(}\ref{summu}) which we rewrite as
\begin{equation}
C_{x,p}=0,\;\;\;\;\;\;\;\mathrm{where}\;\;\;\;\;\;\;\;C_{x,p}=-1+%
\sum_{j=0}^{x}\mu _{x,p,j}.  \label{constr1}
\end{equation}
This is not yet amenable to traditional differential techniques, since the
set $\Bbb{R}_{+}^{N}\cap \mathcal{U}$ is not open.\ However, it is not
difficult to bar out the vectors on the boundary as candidates for global,
indeed even local, maximizers, as we now proceed to do.

Let us compute the gradient of the function of $\mathbf{\mu }$ maximized in (%
\ref{pbmax1}), say $f_{1}\left( \mathbf{\mu }\right) $. For the quantity
inside the braces, using (\ref{defb23}) and (\ref{sumbeta}), and setting
\[
U=\frac{9\left( 1-\phi \right) \beta _{3}}{\left( 3\phi -\beta _{1}\right)
\beta _{2}}\;\;\;\;\;\;\;\;\;\mathrm{and}\;\;\;\;\;\;\;\;\;V=1+\frac{\beta
_{2}^{2}}{3\left( 3\phi -\beta _{1}\right) \beta _{3}}=\frac{\left( \beta
_{1}+6\phi -3\right) ^{2}}{3\beta _{3}\left( 3\phi -\beta _{1}\right) },
\]
we obtain for $\nabla f_{1}:$%
\begin{eqnarray*}
&&3\log \frac{3\phi -\beta _{1}}{3\left( 1-\phi \right) }.\nabla \phi -\log
\left( 3\phi -\beta _{1}\right) .\nabla \beta _{1}-\log \beta _{2}.\nabla
\beta _{2}-\log \left( 3\beta _{3}\right) .\nabla \beta _{3}-\nabla \beta
_{1}-\nabla \beta _{2}-\nabla \beta _{3} \\
&=&-3\log U.\nabla \phi +\log \left( V-1\right) .\nabla \beta _{1},
\end{eqnarray*}
so, taking the $\left( x,p,j\right) $-coordinate:
\begin{equation}
\frac{\partial f_{1}}{\partial \mu _{x,p,j}}=\left\{
\begin{array}{ll}
\widetilde{\kappa }_{x,p}\left[ \log \frac{h_{x,p,j}}{\mu _{x,p,j}}-1+\left(
x-2p\right) \log U+\left( p-j\right) \log \left( V-1\right) \right] , &
\;\;\;\;0\leq j\leq p-1; \\
\widetilde{\kappa }_{x,p}\left[ \log \frac{h_{x,p,j}}{\mu _{x,p,j}}-1+\left(
j-p\right) \log \left( V-1\right) \right] , & \;\;\;\;p\leq j\leq x.
\end{array}
\right.  \label{gradf1}
\end{equation}
With this knowledge, we can establish:

\begin{lemma}
\label{bdryremvl} No feasible vector $\mathbf{\mu }$ (i.e. $\mathbf{\mu \in }%
\Bbb{R}_{+}^{N}\cap \mathcal{U}$ satisfying (\ref{constr1})) having at least
one null coordinate can be a local maximizer for the problem (\ref{pbmax1}).
\end{lemma}

\begin{proof}
Choose $j_{1}$ and $j_{2}$ such that $\mu _{x,p,j_{1}}=0$ and $\mu
_{x,p,j_{2}}\neq 0$, and consider the real-valued function $f_{1}^{*}$
defined on $\left] 0,\mu _{x,p,j_{2}}\right[ $ by $f_{1}^{*}\left( \xi
\right) =f_{1}\left( \mathbf{\mu }_{\xi }\right) $, where $\mathbf{\mu }%
_{\xi }$ differs from $\mathbf{\mu }$ only in the $\left( x,p,j_{1}\right) $
and $\left( x,p,j_{2}\right) $ coordinates, the former being equal to $\xi $
and the latter to $\mu _{x,p,j_{2}}-\xi $. Of course, for sufficiently small
$\xi $, $\mathbf{\mu }_{\xi }$ is still feasible, so it suffices to show
that $0$ is not a local maximum for $f_{1}^{*}$. But, using (\ref{gradf1}):
\begin{eqnarray*}
\frac{1}{\tilde{\kappa}_{x,p}}\frac{\partial f_{1}^{*}}{\partial \xi } &=&%
\frac{1}{\tilde{\kappa}_{x,p}}\left[ \frac{\partial f_{1}}{\partial \mu
_{x,p,j_{1}}}\left( \mathbf{\mu }_{\xi }\right) -\frac{\partial f_{1}}{%
\partial \mu _{x,p,j_{2}}}\left( \mathbf{\mu }_{\xi }\right) \right] \\
&=&\log \frac{h_{x,p,j_{1}}}{\xi }-\log \frac{h_{x,p,j_{2}}}{\mu
_{x,p,j_{2}}-\xi }+R_{x,p,j_{1},j_{2}}\log \left[ U\left( \mathbf{\mu }_{\xi
}\right) \right] +S_{p,j_{1},j_{2}}\log \left[ V\left( \mathbf{\mu }_{\xi
}\right) -1\right] ,
\end{eqnarray*}
where $\left| R_{x,p,j_{1},j_{2}}\right| \leq x_{max}$ and $\left|
S_{p,j_{1},j_{2}}\right| \leq x_{max}$. As $\xi \rightarrow 0+$, the first
term on the right tends to $+\infty $, the second remains bounded, while,
since $\mathbf{\mu }_{\xi }$ is feasible,
\[
\frac{9\left( 1-\phi _{max}\right) \beta _{3min}}{\left( 3\phi _{max}-\beta
_{1min}\right) \beta _{2max}}\leq U\left( \mathbf{\mu }_{\xi }\right) \leq
\frac{9\left( 1-\phi _{min}\right) \beta _{3max}}{\left( 3\phi _{min}-\beta
_{1max}\right) \beta _{2min}}
\]
and
\[
\frac{\beta _{2min}^{2}}{3\left( 3\phi _{max}-\beta _{1min}\right) \beta
_{3max}}\leq V\left( \mathbf{\mu }_{\xi }\right) -1\leq \frac{\beta
_{2max}^{2}}{3\left( 3\phi _{min}-\beta _{1max}\right) \beta _{3min}},
\]
so the third and fourth terms stay finite too. All in all, $\partial
f_{1}^{*}/\partial \xi $ is seen to tend to $+\infty $ as $\xi \rightarrow
0+ $, so obviously (e.g., from the mean value formula) $0$ cannot be a local
maximum for $f_{1}^{*}$.
\end{proof}

So, the set constraint in problem (\ref{pbmax1}) may be replaced by $\mathbf{%
\mu }\in \mathcal{O}$, with $\mathcal{O}=\left] 0,+\infty \right[ ^{N}\cap
\mathcal{U}$ (an \textit{open }subset of $\Bbb{R}^{N}$), allowing a study of
local optima by traditional differential methods. $f_{1}$ is readily seen to
be bounded in $\mathcal{O},$ since for $x\geq 2p$ all $h_{x,p,j}$'s are $%
\leq 2^{x-p-1}$; thus, if some number $M$ bounds from above all local
maximizers in $\mathcal{O}$, then $f_{1}\left( \mathbf{\mu }\right) \leq M$
for any feasible $\mathbf{\mu }$.

Since all constraints are affine, we do not actually need any further
constraint qualification such as linear independence of the gradients,
though this is clearly the case. The classical method of Lagrange
multipliers applies \cite{Lue84}: a necessary condition for optimality of $%
f_{1}$ at some feasible $\mathbf{\mu }^{*}\in \mathcal{O}$ is \textit{%
stationarity} in the sense that there exist real numbers $\Lambda _{x,p}$, $%
0\leq 2p\leq x\leq x_{max}$, such that at $\mathbf{\mu }^{*}$:
\begin{equation}
\nabla f_{1}+\sum_{0\leq 2p\leq x\leq x_{max}}\Lambda _{x,p}\nabla C_{x,p}=0.
\label{lagr1}
\end{equation}
Now take the $\left( x,p,j\right) $-coordinate of (\ref{lagr1}), using (\ref
{gradf1}): since $\widetilde{\kappa }_{x,p}\neq 0$ whenever $2p\leq x$, we
get, for $0\leq j\leq p-1$ and $p\leq j\leq x$ respectively:
\begin{equation}
\mu _{x,p,j}=h_{x,p,j}\exp \left( \frac{\Lambda _{x,p}}{\tilde{\kappa}_{x,p}}%
-1\right) U^{x-2p}\left( V-1\right) ^{p-j}\;\;\;\;\mathrm{and}\;\;\;\;\mu
_{x,p,j}=h_{x,p,j}\exp \left( \frac{\Lambda _{x,p}}{\tilde{\kappa}_{x,p}}%
-1\right) \left( V-1\right) ^{j-p}.  \label{mustat}
\end{equation}
The Lagrange multipliers are determined by plugging this back into the
constraints (\ref{constr1}):
\begin{eqnarray*}
1 &=&\exp \left( \frac{\Lambda _{x,p}}{\tilde{\kappa}_{x,p}}-1\right) \left[
U^{x-2p}\sum_{j=0}^{p-1}{{{{{{{\binom{p }{j}}}}}}}}\left( V-1\right)
^{p-j}+\sum_{j=p}^{x}{{{{{{{\binom{x-p }{j-p}}}}}}}}\left( V-1\right)
^{j-p}\right] \\
&=&\exp \left( \frac{\Lambda _{x,p}}{\tilde{\kappa}_{x,p}}-1\right) \left[
U^{x-2p}\left( V^{p}-1\right) +V^{x-p}\right] .
\end{eqnarray*}
Hence a necessary (and sufficient) condition for stationarity is that the $%
N\left( x_{max}\right) $ unknowns $\mu _{x,p,j},0\leq 2p\leq x\leq x_{max}$
such that $0\leq j\leq x,$ satisfy the following system of $N\left(
x_{max}\right) $ equations:
\begin{equation}
\mu _{x,p,j}=\left\{
\begin{array}{lll}
{{{{{{{\binom{p }{j}}}}}}}}\frac{U^{x-2p}\left( V-1\right) ^{p-j}}{%
U^{x-2p}\left( V^{p}-1\right) +V^{x-p}} & \mathrm{if} & 0\leq j\leq p-1; \\
&  &  \\
{{{{{{{\binom{x-p }{j-p}}}}}}}}\frac{\left( V-1\right) ^{j-p}}{%
U^{x-2p}\left( V^{p}-1\right) +V^{x-p}} & \mathrm{if} & p\leq j\leq x
\end{array}
\right.  \label{eqmu1}
\end{equation}
(where of course $U$ and $V\;$are themselves fairly complicated functions of
the $\mu $'s). Note also that for any solution $\mathbf{\mu }$, the quantity
$\alpha _{x,p}=\sum_{j=0}^{p-1}\mu _{x,p,j}$ has the summation-free
expression:
\[
\alpha _{x,p}=\frac{U^{x-2p}\left( V^{p}-1\right) }{U^{x-2p}\left(
V^{p}-1\right) +V^{x-p}}=1-\frac{1}{1+\left( \frac{U}{V}\right)
^{x-2p}\left( 1-\frac{1}{V^{p}}\right) }
\]
and systematically equals $0$ for $p=0$. Further, under the same conditions
the coefficient of $\displaystyle{\frac{\widetilde{\kappa }_{x,p}}{c}}$ in $%
\beta _{1}$ has the value
\begin{eqnarray*}
&&\frac{1}{U^{x-2p}\left( V^{p}-1\right) +V^{x-p}}\left[
U^{x-2p}\sum_{h=1}^{p}h{{{{{{{\binom{p }{h}}}}}}}}\left( V-1\right)
^{h}+\sum_{l=0}^{x-p}l{{{{{{{\binom{x-p }{l}}}}}}}}\left( V-1\right)
^{l}\right] \\
&=&\frac{V-1}{V}\frac{pU^{x-2p}V^{p}+\left( x-p\right) V^{x-p}}{%
U^{x-2p}\left( V^{p}-1\right) +V^{x-p}},
\end{eqnarray*}
where again the summation in $j$ has disappeared. The system (\ref{eqmu1})
may seem hopeless at first sight. However, all unknowns can be extracted in
terms of just two (affine) functions of themselves, $\phi $ and $\beta _{1}$%
. This has the following consequence.\ Consider the system $S^{*}$ of $%
N\left( x_{max}\right) +2$ equations in as many unknowns obtained by viewing
$\phi $ and $\beta _{1}$ as two further unknowns, and (\ref{defphimod}) and (%
\ref{defgamma1}) as two additional equations. A solution of (\ref{eqmu1})
immediately gives one of $S^{*},$ and conversely. Solving (\ref{eqmu1})
amounts to solving $S^{*}$ by trivially eliminating $\phi $ and $\beta _{1}.$
But the property just stated means that there is a better way to solve $%
S^{*},$ namely eliminating the $\mu $'s, leaving just $2$ equations in $2$
unknowns. So, viewing now $U$ and $V$ as functions of $\phi $ and $\beta
_{1} $ only, we plug the r.h.s.'s of (\ref{eqmu1}) into (\ref{defphimod}), (%
\ref{defgamma1}) to obtain
\begin{equation}
\lambda \phi =\widetilde{K}-\sum_{1\leq 2p\leq x\leq x_{max}}\widetilde{%
\kappa }_{x,p}\left( x-2p\right) \left[ 1-\frac{1}{1+\left( \frac{U}{V}%
\right) ^{x-2p}\left( 1-V^{-p}\right) }\right] ,  \label{eqphi1}
\end{equation}
\begin{equation}
\beta _{1}c=\frac{V-1}{V}\sum_{0\leq 2p\leq x\leq x_{max}}\widetilde{\kappa }%
_{x,p}\frac{pU^{x-2p}V^{p}+\left( x-p\right) V^{x-p}}{U^{x-2p}\left(
V^{p}-1\right) +V^{x-p}}.  \label{eqb1}
\end{equation}
Having solved this in $\phi $ and $\beta _{1}$, we plug them into (\ref
{eqmu1}) and obtain the $\mu $'s. While still highly nonlinear, the system (%
\ref{eqphi1}, \ref{eqb1}) can, as we shall show, be rigorously analyzed. But
what we certainly cannot do is to exploit convexity considerations as in
\cite{DubBou97}, \cite{BouDub99}: here the objective function is \textit{not
}concave.

\section{The equations: analysis and numerical resolution.}

\subsection{Preliminary transformations.}

In the sequel, $\left( \phi ,\beta _{1}\right) $ will denote an \textit{%
arbitrary} solution of (\ref{eqphi1}, \ref{eqb1}) or an equivalent system.
Before we proceed, it will be helpful to rearrange some of the already
obtained expressions in a more convenient form.

\subsubsection{The expectation revisited.}

Assume that we have a solution $\left( \phi ,\beta _{1}\right) $ to the
system (\ref{eqphi1}, \ref{eqb1}) and that the corresponding parameters $%
\mathbf{\mu }$ do give rise to the (global) maximum appearing in (\ref{tfb1}%
). We show that our bound on the expectation simplifies to a formula where
the parameters $\mathbf{\mu }$ only intervene via the two quantities $\phi $
and $\beta _{1}$ (and $U$ and $V$ viewed as functions of these).

First we have, in view of (\ref{eqmu1}) and taking into account \textit{(}%
\ref{summu}), (\ref{defgamma1}), (\ref{alphadef}), and (\ref{defphimod}):
\[
\prod\limits_{0\leq 2p\leq x\leq x_{max}}\left[ \prod\limits_{j=0}^{x}\left(
\frac{\mu _{x,p,j}}{h_{x,p,j}}\right) ^{\mu _{x,p,j}}\right] ^{\widetilde{%
\kappa }_{x,p}}=\frac{U^{\widetilde{K}-\lambda \phi }\left( V-1\right)
^{\beta _{1}c}}{\prod\limits_{0\leq 2p\leq x\leq x_{max}}\left[
U^{x-2p}\left( V^{p}-1\right) +V^{x-p}\right] ^{\widetilde{\kappa }_{x,p}}}.
\]
Call the inverse of the r.h.s. $g_{2}\left( \phi ,\beta _{1}\right) $, plug
it back into (\ref{tfb1}), and modify (\ref{firstEbound}) accordingly, using
$\left( 2\varepsilon n\right) ^{D/n}\leq \exp \left( 2\varepsilon D/e\right)
:$%
\begin{equation}
\begin{array}{lll}
\mathbf{E}\left( X_{n,\varepsilon ,x_{max},c}\right) ^{1/n} & \leq & \left(
6n^{3}\right) ^{\frac{1}{n}}3^{c}\left( \frac{\lambda }{6e}\right) ^{\lambda
}\displaystyle{\frac{2^{\sum_{x>2p}\widetilde{\kappa }_{x,p}}\exp \left(
2\varepsilon D/e\right) }{\prod\limits_{0\leq 2p\leq x\leq x_{max}}\left[
p!\left( x-p\right) !\tilde{\kappa}_{x,p}\right] ^{\widetilde{\kappa }_{x,p}}%
}}G_{1}\left( \varepsilon ,x_{max}\right) \times \\
&  & G_{2}\left( \varepsilon ,x_{max}\right) g_{2}\left( \phi ,\beta
_{1}\right) \left[ \displaystyle{\frac{\left( 3\phi -\beta _{1}\right)
^{3\phi -\beta _{1}}\left[ 3\left( 1-\phi \right) \right] ^{3\left( 1-\phi
\right) }}{\beta _{2}^{\beta _{2}}\left( 3\beta _{3}\right) ^{\beta _{3}}}}%
\right] ^{c}.
\end{array}
\label{practicalbound}
\end{equation}
This is essentially the estimate that will serve in our numerical
evaluations. It is possible further to transform it so that all exponents
become fixed (i.e. independent of $\phi $ and $\beta _{1}$), but this,
although noteworthy, will not be used here.

Let us emphasize that the function of $\phi $ and $\beta _{1}$ on the right
of (\ref{practicalbound}) has little to do with the objective function in (%
\ref{tfb1}) (a function of $\mathbf{\mu }$, anyway).\ All we say is that it
dominates $\mathbf{E}\left( X_{n,\varepsilon ,x_{max},c}\right) ^{1/n}$ for
\textit{some} pair(s) $\left( \phi ,\beta _{1}\right) $ satisfying the
system (\ref{eqphi1}, \ref{eqb1}), and our final bound will be valid for
\textit{any} such solution, without having to assume or prove uniqueness.
Although it can be seen that (\ref{eqphi1}, \ref{eqb1}) actually
characterizes stationary values of that function too, the (in fact unique)
solution is not a maximum but a saddle point.

\subsubsection{A modified form of the second equation.}

The numerator of the fraction in the sum on the right-hand side of (\ref
{eqb1}) can be written
\[
pU^{x-2p}V^{p}+\left( x-p\right) V^{x-p}=\left( 2p-x\right) U^{x-2p}\left(
V^{p}-1\right) +\left( x-p\right) \left[ U^{x-2p}\left( V^{p}-1\right)
+V^{x-p}\right] +pU^{x-2p},
\]
so that (\ref{eqb1}) also reads
\begin{eqnarray*}
\beta _{1}c &=&\frac{V-1}{V}\left[ -\sum_{0\leq 2p\leq x\leq
x_{max}}H_{x,p}\alpha _{x,p}+\sum_{0\leq 2p\leq x\leq x_{max}}\left(
x-p\right) \widetilde{\kappa }_{x,p}\right. \\
&&\left. +\sum_{2\leq 2p\leq x\leq x_{max}}p\widetilde{\kappa }_{x,p}\frac{%
U^{x-2p}}{U^{x-2p}\left( V^{p}-1\right) +V^{x-p}}\right] .
\end{eqnarray*}
The second sum is by definition $\widetilde{K}$, while if (\ref{eqphi1}) is
verified, the first is $\widetilde{K}-\lambda \phi $. Thus the system (\ref
{eqphi1},\ref{eqb1}) is equivalent to (\ref{eqphi1},\ref{eqb1mod}), where (%
\ref{eqb1mod}) is as follows:
\begin{equation}
\beta _{1}c=\frac{V-1}{V}\left[ \lambda \phi +\sum_{2\leq 2p\leq x\leq
x_{max}}\widetilde{\kappa }_{x,p}\frac{pU^{x-2p}}{U^{x-2p}\left(
V^{p}-1\right) +V^{x-p}}\right] .  \label{eqb1mod}
\end{equation}

\subsubsection{The monotone behaviour of $U$ and $V$ in each variable
separately.}

Set [viewing $\beta _{2},\beta _{3}$ as functions $\widetilde{\beta }_{2},%
\widetilde{\beta }_{3}$ of $\phi ,\beta _{1},$ cf. (\ref{defb23})] $\mathcal{%
D}_{\phi ,\beta _{1}}=\left[ \phi _{min},\phi _{max}\right] \times \left[
\beta _{1min},\beta _{1max}\right] \cap \widetilde{\beta }_{2}^{-1}\left(
\left[ \beta _{2min},\beta _{2max}\right] \right) \cap \widetilde{\beta }%
_{3}^{-1}\left( \left[ \beta _{3min},\beta _{3max}\right] \right) .\;$Within
our range of $c,$ we can disregard pairs $\left( \phi ,\beta _{1}\right)
\notin \mathcal{D}_{\phi ,\beta _{1}},$ so we will limit our study of (\ref
{eqphi1},\ref{eqb1mod}) to $\mathcal{D}_{\phi ,\beta _{1}}.$

For fixed $\phi $ and variable $\beta _{1}$ within $\mathcal{D}_{\phi ,\beta
_{1}}$, then, $U$ increases (strictly) as the quotient of an increasing
numerator by a decreasing denominator.

For fixed $\beta _{1}$, $U$ increases (strictly) in $\phi $ since $\beta
_{1max}<1$ implies
\[
\frac{\partial \log U}{\partial \phi }=\frac{-1}{1-\phi }+\frac{3}{\beta _{3}%
}-\frac{3}{3\phi -\beta _{1}}+\frac{3}{\beta _{2}}=\frac{2\beta _{1}}{\left(
1-\phi \right) \beta _{2}}+\frac{6\left( 1-\beta _{1}\right) }{\beta
_{3}\left( 3\phi -\beta _{1}\right) }>0
\]
(As an unconstrained linear combination of the $\mu _{x,p,j}$'s, $\beta _{1}$
does reach values $>1$.)

For fixed $\phi $, $V-1$ has a decreasing numerator, while the denominator,
three times a product of factors with a constant sum, increases until $\beta
_{1}$ reaches $\beta _{1M}$ such that $\beta _{3}=3\phi -\beta _{1}$;
however, $\beta _{1M}=1>\beta _{1max}$, so $V$ decreases on $\mathcal{D}%
_{\phi ,\beta _{1}}$.

For fixed $\beta _{1}$, $V-1$ also decreases owing to a decreasing
denominator and increasing numerator.

To sum up: with either variable fixed, $U$ increases and $V$ decreases in
the other variable.

\subsubsection{Bounds on $U$ and $V.$}

We henceforth set $c=c_{min}=c_{max}=4.506.\;$All the foregoing remains
valid, but some inequalities become tighter, starting with (\ref
{GamPsiConstr}) and (\ref{BetaPhiConstr}); in the latter we can now take $%
\beta _{1min}=\beta _{2min}=0.33018;\beta _{1max}=\beta
_{2max}=0.52891;\beta _{3min}=0.077639;\beta _{3max}=0.21782;\phi
_{min}=0.525245;$ and $\phi _{max}=0.619063.$ (Also, e.g. now $R_{3}\left(
\varepsilon ,x_{max}\right) $ has a smaller value $<1.104\;10^{-11}.$)

These limitations imply helpful ones on $U$ and $V:\;\;\;U\leq
U_{max1}=2.69268,\;\;U\leq U_{max2}/\left( V-1\right) $ where $%
U_{max2}=0.687424,\;\;V\geq V_{min1}=1.109255,\;\;U/V\geq \left( U/V\right)
_{max1}=11.2022;$ we need and prove better ones than the last two.

We solve the \textit{constrained }minimization problem (with variables $\phi
$ and $\beta _{1}$): minimize $V$, subject to the $8$ linear constraints
written above. These define a convex polygonal domain with $7$ sides in the
plane $\left( \phi ,\beta _{1}\right) $. Due to the decreasing character of $%
V$ in each variable, the minimum can obviously not be attained at an
interior point, nor along any side other than the two given by $\beta
_{3}=\beta _{3max}$ and $\beta _{2}=\beta _{2min}$. $V$ is easily seen to
decrease in $\beta _{1}$ along the first and to increase in $\beta _{1}$
along the second, so the minimum is attained at their intersection, and is
found equal to:
\[
V_{min2}=1+\frac{\beta _{2min}^{2}}{3\beta _{3max}\left( 2\beta
_{2min}+3\beta _{3max}\right) }=1.126983.
\]
We can maximize $\left( U/V\right) $ in a very similar way, since it
increases in each variable separately. Again, the maximum must be along one
of the same two sides of the same polygon, as seen using the form $%
U/V=27\left( 1-\phi \right) \beta _{3}^{2}/\left[ \beta _{2}\left( \beta
_{1}+6\phi -3\right) ^{2}\right] ,$ and equals
\[
\left( \frac{U}{V}\right) _{max2}=\frac{9\left[ 2\left( 1-\beta
_{3max}\right) -\beta _{2min}\right] \beta _{3max}^{2}}{\beta _{2min}\left(
\beta _{2min}+3\beta _{3max}\right) ^{2}}=1.64966.
\]

\subsection{Outline.}

From now on, $c$ is taken to be equal to $4.506$. The remainder of
the paper is devoted to showing that for this $c$, the product
$2^{\left( \rho +\varepsilon \Delta \right) n}\mathbf{E}\left(
X_{n,\varepsilon ,x_{max},c}\right) $ tends to $0$ as
$n\rightarrow \infty $. Since the probability of satisfiability
decreases in $c$, this will establish that the threshold is below
$4.506$.

% POUR WINEDIT ENLEVER LA MISE EN COMMENTAIRE CI-DESSOUS JUSQU'A FIN

\begin{figure}[h]
\begin{center}
\includegraphics[width=100mm,height=60mm]{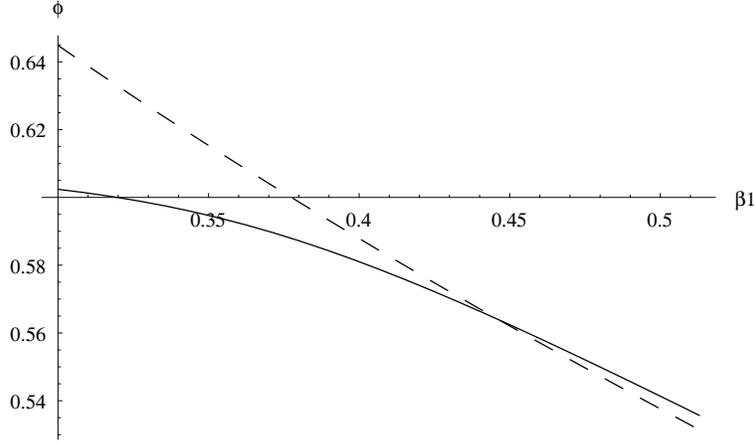}
\hspace{+0cm}\parbox{10cm}{\vspace{1.8cm}\caption{The solutions
$(\beta _{1},\phi)$ of equations (29) (dashed line) and (30)
(solid line)}} \label{fig: toto }
\end{center}
\end{figure}

% POUR WINEDIT FIN ENLEVER LA MISE EN COMMENTAIRE CI-DESSUS

While Figure 1 clearly suggests that the system (\ref{eqphi1}, \ref{eqb1mod}%
) has a unique solution, we present a strictly rigorous analysis.\ It
exploits special features of this system, leading to numerical calculations
which can be reliably and routinely performed to any desired precision.

A close study of the 2 equations, written as $Eq_{1}=0$ and $Eq_{2}=$ $0,$
shows that each defines $\phi $ as a unique decreasing function of $\beta
_{1}$; then a constructive numerical procedure is applied to narrow down the
location of \textit{any} common root. Uniqueness is neither assumed nor
proven, though this could be done with a little more effort.

Actually, it suffices to establish the (strict) monotony of $Eq_{1}$ and $%
Eq_{2}$ in each variable separately; the monotone behavior of the
corresponding implicit functions follows.\ And it turns out that it is
easier to reason directly in terms of this separate monotony, and that in
this approach strictness is not used.

There is a slight restriction to the monotony of $Eq_{1}$, which does not
affect the end result. A precise statement follows.

\begin{proposition}
\label{sepmon} \textit{i)} $Eq_{2}$ decreases strictly in each variable
separately over the whole domain of interest $\mathcal{D}_{\phi ,\beta _{1}}$%
, and even over the wider set $\left[ \phi _{min},\phi _{max}\right] \times
\left[ \beta _{1min},\beta _{1max}\right] $.\newline
\textit{(ii)} For any $\phi \in \left[ \phi _{min},\phi _{max}\right] $
(resp. any $\beta _{1}\in \left[ \beta _{1min},\beta _{1max}\right] $),
there exists $\beta _{1}^{*}\left( \phi \right) \leq 1$ (resp. $\phi
^{*}\left( \beta _{1}\right) \leq 1$) such that $Eq_{1}\left( \phi ,.\right)
<0$ over the interval $\left[ \beta _{1}^{*}\left( \phi \right) ,1\right] $,
(resp. such that $Eq_{1}\left( .,\beta _{1}\right) <0$ over the interval $%
\left[ \phi ^{*}\left( \beta _{1}\right) ,1\right] $) and that $Eq_{1}\left(
\phi ,.\right) $ decreases strictly on $\left[ 0,\beta _{1}^{*}\left( \phi
\right) \right] $ (resp. $Eq_{1}\left( .,\beta _{1}\right) $ decreases
strictly on $\left[ 0,\phi ^{*}\left( \beta _{1}\right) \right] $). In
particular, if $Eq_{1}\left( \phi ,\beta _{1}\right) =0$, then $\beta
_{1}^{*}\left( \phi \right) >\beta _{1}$ and $\phi ^{*}\left( \beta
_{1}\right) >\phi $.
\end{proposition}

The remainder of this section is devoted to proving Proposition \ref{sepmon}%
. We start with the equation that is monotone on the whole of $\mathcal{D}%
_{\phi ,\beta _{1}}.$

\subsection{The second equation, separate monotony.}

We use the modified form (\ref{eqb1mod}), and deal successively with fixed $%
\phi $, variable $\beta _{1},$ and the reverse.\ First, some considerations
which apply to both. (\ref{eqb1mod}) can be rewritten equivalently:
\begin{equation}
0=Eq_{2}\equiv -\beta _{1}c+\lambda \phi \left( 1-\frac{1}{V}\right)
+\sum_{1\leq 2p\leq x\leq x_{max}}p\widetilde{\kappa }_{x,p}\frac{V-1}{%
V\left( V^{p}-1\right) }\left[ 1-\frac{1}{1+\left( \frac{U}{V}\right)
^{x-2p}\left( 1-\frac{1}{V^{p}}\right) }\right] .  \label{eqb1mod2}
\end{equation}
Call the denominator of the last fraction $D_{x,p}$. We differentiate one
variable, leaving the other fixed.\ The derivatives are denoted simply by a
prime because the context will always make the meaning clear. The following
equality holds in either case.
\[
\left\{ \frac{V-1}{V\left( V^{p}-1\right) }\left[ 1-\frac{1}{D_{x,p}}\right]
\right\} ^{\prime }=\frac{-V^{\prime }}{V^{2}}\frac{pV^{p+1}-\left(
p+1\right) V^{p}+1}{\left( V^{p}-1\right) ^{2}}\left[ 1-\frac{1}{D_{x,p}}%
\right] +
\]
\[
\;\;\;\;\;\;\;\;\;\;\;\;\;\;\;\;\;\;\;\;\;\;\;\;\;\;\;\;\;\;\;\;\;\;\;\;\;\;%
\;\frac{V-1}{V\left( V^{p}-1\right) }\frac{\left( \frac{U}{V}\right)
^{x-2p-1}}{D_{x,p}^{2}}\left[ \left( x-2p\right) \left( \frac{U^{\prime }}{U}%
-\frac{V^{\prime }}{V}\right) \left( 1-\frac{1}{V^{p}}\right) +\frac{%
pV^{\prime }}{V^{p+1}}\right] .
\]
It will be shown, in the relevant subsections, that

\begin{lemma}
\label{R} Let $X=3\phi -1$, $Y=3\phi -\beta _{1}$, $Z=Y-X=1-\beta _{1}$. We
have
\[
\frac{U^{\prime }}{U}-\frac{V^{\prime }}{V}=\frac{-V^{\prime }}{V\left(
V-1\right) }R,
\]
where $R$ is a positive quantity such that for fixed $\phi $%
\[
R=\frac{Y}{X}V-1,
\]
while for fixed $\beta _{1}$%
\[
R<S=\frac{Y}{Z}\left( V-1\right) +\frac{1}{2}.
\]
\end{lemma}

It follows that
\begin{eqnarray*}
\left\{ \frac{V-1}{V\left( V^{p}-1\right) }\left[ 1-\frac{1}{D_{x,p}}\right]
\right\} ^{\prime } &=&\frac{-V^{\prime }}{V^{2}}\left\{ \frac{%
pV^{p+1}-\left( p+1\right) V^{p}+1}{\left( V^{p}-1\right) ^{2}}\left[ 1-%
\frac{1}{D_{x,p}}\right] \right. \\
&&\left. +\frac{\left( \frac{U}{V}\right) ^{x-2p-1}}{V^{p}D_{x,p}^{2}}\left[
\left( x-2p\right) R-p\frac{V-1}{V^{p}-1}\right] \right\} .
\end{eqnarray*}
The two terms inside the curly brackets on the right will be called $A_{x,p}$
and $B_{x,p}$, respectively. We need to study the fraction in $V$ that
occurs in $A_{x,p}$:

\begin{lemma}
\label{Vfrac} For nonzero $p$, the quantity $\displaystyle{\frac{%
pV^{p+1}-\left( p+1\right) V^{p}+1}{\left( V^{p}-1\right) ^{2}}}$ decreases
in $V>1$ (and tends to $\frac{1+1/p}{2}$ as $V\rightarrow 1+0$).
\end{lemma}

A\ standard exercise in derivatives and infinitesimals.

\subsubsection{Fixed $\phi $, variable $\beta _{1}$.}

\textbf{Proof of Lemma \ref{R}} (fixed $\phi $). Note that $\beta
_{2}=-3X+2Y $, $\beta _{3}=2X-Y$; also, $V^{\prime }/V=\frac{V-1}{V\beta
_{2}\beta _{3}Y}2X\left( Y-3X\right) $ and $U^{\prime }/U=\frac{V}{V-1}%
\left( 1-\frac{Y}{X}\right) \frac{V^{\prime }}{V},$ so
\[
\frac{U^{\prime }}{U}-\frac{V^{\prime }}{V}=\frac{V^{\prime }}{V\left(
V-1\right) }\left[ \left( V-\frac{Y}{X}V\right) -\left( V-1\right) \right] .
\]
That $R$ is positive results from the fact that $\left( U/V\right) $ is
increasing and $V$ decreasing (in $\beta _{1}$).

We now show that $R$ can be expressed as a function of $V$\ alone.

\begin{proposition}
\label{RVphi} For constant $\phi $,
\[
R=\frac{3V^{2}-1+3V\sqrt{V\left( V-1\right) }}{3V+1},
\]
and this function is concave in $V>1$.
\end{proposition}

\begin{proof}
Indeed, $Y/X=\left( R+1\right) /V$ and $V-1=\left( 2\frac{Y}{X}-3\right)
^{2}/\left[ 3\left( 2-\frac{Y}{X}\right) \frac{Y}{X}\right] ,$ whence a
second-degree relationship between $V$ and $R+1$ which can be solved in $%
R+1: $%
\[
R+1=\frac{3V}{3V+1}\left[ V+1+\omega \sqrt{V\left( V-1\right) }\right]
\]
where $\omega =\pm 1$. The coefficient of $\omega $ in $R-1/2$ is larger
than the $\omega $-free term, while from the definition and (\ref{defb23}), $%
R-1/2$ is seen to equal $1/6\;Y/X\;\beta _{2}/\beta _{3}$ which is positive
on $\mathcal{D}_{\phi ,\beta _{1}}.$ Therefore, $\omega $ must be $+1.$ As
for concavity, $R^{\prime \prime }$ is found to have the sign of $%
11V^{2}-30V+3-16\left( V-1\right) \sqrt{V\left( V-1\right) }$, or, in terms
of $W=V-1$:
\[
11W^{2}-8W-16-16W\sqrt{W\left( W+1\right) }<-5W^{2}-8W-16<0.
\]
\end{proof}

\begin{corollary}
\label{affboundphi} We have the following affine upper bound for $R$:
whenever $V\geq V_{min2}$, and irrespective of the constant value of $\phi $%
,
\[
R\leq a_{1}V+b_{1},
\]
where $a_{1}=2.4427$ and $b_{1}=-1.8194$.
\end{corollary}

\begin{proof}
The curve is below its tangent at any point, but since we need better and
better estimates as $V$ decreases, it is appropriate to choose the tangent
at $V_{min2}$. This gives the stated values of the coefficients.
\end{proof}

Since $-V^{\prime }/V^{2}>0$, in order to prove that \ref{eqb1mod2}
decreases in $\beta _{1}$ it (amply) suffices to show that
\[
\lambda \phi _{min}\geq A+B,\;\mathrm{where}\;A=\sum_{2\leq 2p\leq x\leq
x_{max}}p\widetilde{\kappa }_{x,p}A_{x,p}\;\mathrm{and}\;B=\sum_{3\leq
2p+1\leq x\leq x_{max}}p\widetilde{\kappa }_{x,p}B_{x,p},
\]
and $\lambda \phi _{min}=3\times 4.506\times 0.525245>7.1$.

For $A$, we use Lemma \ref{Vfrac}:
\[
A\leq \sum_{2\leq 2p\leq x\leq x_{max}}p\tilde{\kappa}_{x,p}\frac{%
pV_{min2}^{p+1}-\left( p+1\right) V_{min2}^{p}+1}{\left(
V_{min2}^{p}-1\right) ^{2}}\left[ 1-\frac{1}{1+\left( \frac{U}{V}\right)
_{max2}^{x-2p}}\right] ,
\]
or less than $1.894.$ As regards $B$, we keep only the positive terms. Using
Corollary \ref{affboundphi} and the inequality $\frac{r}{\left( 1+rs\right)
^{2}}\leq \frac{1}{4s}$ which is valid for any pair $\left( r,s\right) $ of
positive reals :
\[
B\leq \sum_{3\leq 2p+1\leq x\leq x_{max}}p\left( x-2p\right) \widetilde{%
\kappa }_{x,p}\frac{aV-\left| b\right| }{4\left( V^{p}-1\right) }.
\]
But,
\[
\frac{aV-\left| b\right| }{\left( V^{p}-1\right) }=\left| b\right| \frac{%
\frac{a}{\left| b\right| }V-1}{V-1}\frac{1}{1+V+V^{2}+...+V^{p-1}}.
\]
Since $a>\left| b\right| $, the homographic fraction on the r.h.s. decreases
in $V>1$, as does the last fraction. So, our bound for $B$ can be made
independent of $V\geq V_{min2}$ by evaluating it at $V_{min2}$. This yields $%
B\leq 4.2269$, so the sum $A+B$ is less than $6.125$. This closes the case
of fixed $\phi $, variable $\beta _{1}$.

\subsubsection{Fixed $\beta _{1}$, variable $\phi $.}

\textbf{Proof of Lemma \ref{R}} (fixed $\beta _{1}$). Here we use $Y$ and $Z$%
. Observing that $\beta _{2}=-Y+3Z$, $\beta _{3}=Y-2Z$, we find $V^{\prime
}/V=6\frac{V-1}{V\beta _{2}\beta _{3}Y}Z\left( 3Z-2Y\right) $ and $U^{\prime
}/U=-\frac{1}{1-\phi }+\frac{3Y}{\beta _{2}\beta _{3}}+\frac{V}{V-1}\frac{%
V^{\prime }}{V}$, so that
\[
R=\frac{V\left( V-1\right) }{V^{\prime }\left( 1-\phi \right) }-\frac{%
3YV\left( V-1\right) }{\beta _{2}\beta _{3}V^{\prime }}-1.
\]
As before, $R$ must be positive, and since the first term is negative, it
suffices to show that the sum of the remaining two has the expression stated
above for $S$. Remarking that $V-1=\frac{\beta _{2}}{2}\left( \frac{1}{%
3\beta _{3}}-\frac{1}{3\phi -\beta _{1}}\right) $ and also $V=\frac{\left(
2Y-3Z\right) ^{2}}{3Y\left( Y-2Z\right) }$, we obtain from the expression of
$V^{\prime }/V:$%
\[
-\frac{3YV\left( V-1\right) }{\beta _{2}\beta _{3}V^{\prime }}-1=\frac{Y^{2}V%
}{2Z\left( 2Y-3Z\right) }-1=\frac{1}{2}+\frac{Y}{Z}\frac{3Z-Y}{2}\left[
\frac{-1}{Y}+\frac{1}{3\left( Y-2Z\right) }\right] ,
\]
and conclude using (\ref{defb23}).

We now express $S$ in terms of $V$ alone:

\begin{proposition}
\label{SVbeta} For $V>1$ and $V\neq 4/3$,
\[
S=\frac{1}{2}+\frac{3\left( V-1\right) }{3V-4}\left[ V-2+\sqrt{V\left(
V-1\right) }\right] ,
\]
a concave function of $V>1$ which does not actually have a singularity at $%
V=4/3$.
\end{proposition}

\begin{proof}
Since $Y/Z=\left( S-1/2\right) /\left( V-1\right) $ and $V=\left( 3-2\frac{Y%
}{Z}\right) ^{2}/\left[ 3\frac{Y}{Z}\left( \frac{Y}{Z}-2\right) \right] ,$
we have a relationship between $S_{1}=\left( S-\frac{1}{2}\right) $ and $%
W=\left( V-1\right) $ which is quadratic in each separately, $0=\left(
3W-1\right) S_{1}^{2}-6W\left( W-1\right) S_{1}-9W^{2}$, which we solve in $%
S_{1}$:
\[
S_{1}=\frac{3W}{3W-1}\left[ W-1+\omega \sqrt{W\left( W+1\right) }\right] .
\]
However, since $Z$ is constant, $S$ cannot present a singularity at $W=1/3$,
which rules out $\omega =-1$ in that vicinity, and by continuity for all $%
W.\;$(We can also, for $W\neq 1/3,$ derive straight contradictions from $%
\omega =-1,$ as done in the fixed $\phi $ case.)\newline
As to concavity, for $0<W\neq 1/3$, we easily check that
\[
\frac{d^{2}S}{dV^{2}}=\frac{-3\left( 5W+9\right) }{4\left( W+1\right) \left[
16W\left( W+1\right) ^{2}+\left( 11W^{2}+30W+3\right) \sqrt{W\left(
W+1\right) }\right] }<0;
\]
for $W=1/3,$ this second derivative extends by continuity, hence $%
d^{2}S/dV^{2}$ also exists and is negative there.
\end{proof}

\begin{corollary}
\label{affboundbeta} We have the following affine upper bound for $R$:
whenever $V\geq V_{min2}$, and irrespective of the constant value of $\beta
_{1}$,
\[
R\leq a_{1}V+b_{1},
\]
where $a_{1}=2.2377$ and $b_{1}=-1.7173$.
\end{corollary}

\begin{proof}
The coefficients are those of the tangent to $S$ at $V=V_{min2}$. Note that
again, we have $a_{1}>\left| b_{1}\right| $.
\end{proof}

Proving that (\ref{eqb1mod2}) decreases in $\phi $ now boils down to showing
that, with similar notations to the above:
\begin{equation}
\lambda \phi _{min}\geq -\lambda \frac{V\left( V-1\right) }{V^{\prime }}%
+A+B,\;\mathrm{where\;}A=\sum_{2\leq 2p\leq x\leq x_{max}}p\widetilde{\kappa
}_{x,p}A_{x,p}\;\mathrm{and\;}B=\sum_{3\leq 2p+1\leq x\leq x_{max}}p%
\widetilde{\kappa }_{x,p}B_{x,p}.  \label{NegDeq2/Dphi}
\end{equation}
(of course, all derivatives are now understood to be in $\phi $ for constant
$\beta _{1}$.)

For $A$, we again have the bound $1.894$. For $B$, the same estimate again
applies, \textit{mutatis mutandis, }i.e. with $a_{1}$ and $b_{1}$ replacing $%
a$ and $b$ respectively.\ This gives
\[
B\leq \sum_{3\leq 2p+1\leq x\leq x_{max}}p\left( x-2p\right) \widetilde{%
\kappa }_{x,p}\frac{a_{1}V_{min2}-\left| b_{1}\right| }{4\left(
V_{min2}^{p}-1\right) }<3.643.
\]
And finally, from the expression of $V^{\prime }/V$ and $V=\left( \beta
_{1}+6\phi -3\right) ^{2}/\left( 3\beta _{3}Y\right) $:
\[
-\frac{V\left( V-1\right) }{V^{\prime }}=\frac{\beta _{2}\left( \beta
_{1}+6\phi -3\right) }{18\left( 1-\beta _{1}\right) }=\frac{1}{36\left(
1-\beta _{1}\right) }\left( 6-6\phi -4\beta _{1}\right) \left( \beta
_{1}+6\phi -3\right)
\]
which is maximized by equating the two factors on the right (with a constant
sum), so that
\[
-\lambda \frac{V\left( V-1\right) }{V^{\prime }}\leq \frac{\lambda \left(
1-\beta _{1min}\right) }{16}<0.566.
\]
Bringing all this together, irrespective of the constant value of $\beta
_{1} $, for the right-hand side of (\ref{NegDeq2/Dphi}) we obtain the bound $%
1.894+3.643+0.566=6.103$, which is indeed less than $7.1$.

\subsection{The first equation, separate monotony.}

Actually,as already stated, monotony does not always hold on the whole of $%
\mathcal{D}_{\phi ,\beta _{1}}$ for the first equation
\begin{equation}
0=Eq_{1}\equiv \widetilde{K}-\lambda \phi -\sum_{2\leq 2p<x\leq x_{max}}%
\widetilde{\kappa }_{x,p}\left( x-2p\right) \left[ 1-\frac{1}{1+\left( \frac{%
U}{V}\right) ^{x-2p}\left( 1-V^{-p}\right) }\right] ,  \label{eqphi2}
\end{equation}
at least not in $\beta _{1},$ nor is it strictly needed in order reliably to
locate any solution of the system.\ We shall only prove that, with one
variable kept fixed :

\textbf{Claim 1. }$Eq_{1}$ \textit{decreases from a positive to a negative
value, then stays negative;}\newline
\textit{i.e., the region where monotony may fail contains no solutions
anyway. We write the fraction in }(\ref{eqphi2}) \textit{as} $1/\left(
1+E_{x,p}\right) $.

\subsubsection{Fixed $\phi $, variable $\beta _{1}$.}

\begin{lemma}
\label{monotb1(x>2p)} $E_{x,p}$ increases in $\beta _{1}$ for $x\geq 2p+2$,
while for $x=2p+1$ it increases at least for $0\leq \beta _{1}\leq \beta
_{1}^{*}$ where $\beta _{1}^{*}=\sqrt{3}-3\left( \sqrt{3}-1\right) \phi $.
\end{lemma}

\begin{proof}
Note that $\phi _{min}>1/3$ and $\sqrt{3}-3\left( \sqrt{3}-1\right) \phi
_{max}>0$, so that $\beta _{1}^{*}$ is indeed between $0$ and $1$. It is
readily seen that $E_{4,1}=\left[ 27\left( 1-\phi \right) \beta
_{3}^{2}/\left( \beta _{1}+6\phi -3\right) ^{3}\right] ^{2},$ so that (with $%
X$, $Y$, $Z$ as before) $\partial E_{4,1}/\partial \beta _{1}$ has the sign
of $2/\beta _{3}-3/\left( \beta _{1}+6\phi -3\right) ,$ or of $2\left(
2Y-3Z\right) -3\left( Y-2Z\right) =Y>0.$ Therefore, for $x\geq 2p+2$,
recalling that $U$ increases and $V$ decreases, $E_{x,p},$ which for $p\geq
1 $ equals $\left( \frac{U}{V}\right) ^{x-2p-2}E_{4,1}\left( 1+\frac{1}{V}+%
\frac{1}{V^{2}}+...+\frac{1}{V^{p-1}}\right) ,$ clearly increases.\ For $%
x=2p+1$, the derivative $\partial E_{3,1}/\partial \beta _{1}$ of $%
E_{3,1}=27\left( 1-\phi \right) \beta _{2}\beta _{3}^{2}/\left( \beta
_{1}+6\phi -3\right) ^{4}$ has the sign of $-2/\beta _{2}+2/\beta
_{3}-4/\left( \beta _{1}+6\phi -3\right) $, or of $\left(
Y^{2}/X^{2}-3\right) .$ As $\beta _{1}$ increases from $0$ to $1$, $Y/X$
linearly decreases from $3\phi /\left( 3\phi -1\right) $ to $1$, passing
through $\sqrt{3}$ for $\beta _{1}=\beta _{1}^{*}$. Therefore, $E_{3,1}$
increases for $0\leq \beta _{1}\leq \beta _{1}^{*}$, then decreases; and we
see that for $x=2p+1,\;\;E_{x,p}=E_{3,1}\left( 1+\frac{1}{V}+\frac{1}{V^{2}}%
+...+\frac{1}{V^{p-1}}\right) $ of necessity increases in $\beta _{1}$ for $%
0\leq \beta _{1}\leq \beta _{1}^{*}$.
\end{proof}

Now consider $Eq_{1}$ deprived from the terms such that $x=2p+1$; we call
this $Eq_{1}^{*}$. Obviously $Eq_{1}<Eq_{1}^{*}$, and from Lemma \ref
{monotb1(x>2p)} $Eq_{1}^{*}$ decreases for $0\leq \beta _{1}\leq 1$. Thus,
to prove Claim 1 it suffices to show that $Eq_{1}^{*}$ is negative at $\beta
_{1}^{*}$. However, at $\beta _{1}^{*}$ we have $U=3\left( 1-\phi \right)
/\left( 3\phi -1\right) ,\;\;V=2/\sqrt{3,}$ so that $U/V$ decreases in $\phi
>1/3$. Hence at $\beta _{1}^{*}$, whatever the fixed value of $\phi \in
\left[ \phi _{min},\phi _{max}\right] $:
\[
Eq_{1}^{*}\leq \widetilde{K}-\lambda \phi _{min}-\sum_{4\leq 2p+2\leq x\leq
x_{max}}\widetilde{\kappa }_{x,p}\left( x-2p\right) \left[ 1-\frac{1}{%
1+\left( \frac{3\sqrt{3}}{2}\frac{1-\phi _{max}}{3\phi _{max}-1}\right)
^{x-2p}\left( 1-\left( \frac{\sqrt{3}}{2}\right) ^{p}\right) }\right]
\]
which is less than $-0.157<0.$

\subsubsection{Fixed $\beta _{1}$, variable $\phi $.}

\begin{lemma}
\label{monotphi(x>2p)} $E_{x,p}$ increases in $\phi $ for $x\geq 2p+3$
(provided $\beta _{1}\geq \beta _{1min}$), while for $2p+1\leq x\leq 2p+2$
it increases at least for $0\leq \phi \leq \phi ^{*}$, where
\[
\phi ^{*}=\frac{1}{12}\left[ 15-\frac{9}{2-\beta _{1}}-2\beta _{1}-\frac{%
\sqrt{3}\left( 1-\beta _{1}\right) \sqrt{4\beta _{1}^{2}+4\beta _{1}+3}}{%
2-\beta _{1}}\right] .
\]
This value decreases from $3/4$ to $1/3$ as $\beta _{1}$ increases from $0$
to $1$.
\end{lemma}

\begin{proof}
As before, $E_{x,p}$ increases whenever $E_{x-2p+2,1}$ does, because $%
E_{x,p}=E_{x-2p+2,1}\times $\newline
$\left( 1+\frac{1}{V}+\frac{1}{V^{2}}+...+\frac{1}{V^{p-1}}\right) $, and
since $U/V$ is increasing, $E_{x,p}$ increases whenever $E_{x_{0},p}$ does
for some $x_{0}\leq x$. Therefore all we have to show is that $E_{5,1}$
increases, and that for $0\leq \phi \leq \phi ^{*}$ so does $E_{3,1}$.
Regarding the former, $E_{5,1}=3^{9}\left( 1-\phi \right) ^{3}\beta
_{3}^{6}/\left[ \beta _{2}\left( \beta _{1}+6\phi -3\right) ^{8}\right] $
has a derivative $\partial E_{5,1}/\partial \phi $ with the same sign as $%
-3/\left( 1-\phi \right) +3/\beta _{2}+18/\beta _{3}-48/\left( \beta
_{1}+6\phi -3\right) $, or as\ $2X^{2}-7ZX+Z^{2}\left( 7-Z\right) .$
However, the discriminant $Z^{2}\left( 8Z-7\right) =\left( 1-\beta
_{2}\right) ^{2}\left( 1-8\beta _{1}\right) $ of this quadratic function of $%
X$ remains negative so long as $\beta _{1}\geq \beta _{1min}$, hence the
required monotony.\ Coming now to $\partial E_{3,1}/\partial \phi $, it has
the sign of $-1/\left( 1-\phi \right) -3/\beta _{2}+6/\beta _{3}-24/\left(
\beta _{1}+6\phi -3\right) $, or of $2\left( 1+Z\right) X^{2}-Z\left(
11+2Z\right) X+Z^{2}\left( 11-Z\right) $. This equals zero for
\begin{equation}
X=\frac{Z}{4}\left[ 2+\frac{1}{Z+1}\left( 9+\omega \sqrt{3}\sqrt{%
4Z^{2}-12Z+11}\right) \right]  \label{statE31/phi}
\end{equation}
where $\omega =\pm 1$; however, recalling that $\phi $ cannot exceed $%
1-2\beta _{1}/3=\left( 1+2Z\right) /3$ lest $\beta _{2}$ should become
negative, we see that $X-2Z$ cannot be $>0,$ while if $\omega $ were $+1,$ $%
X-2Z$ would be the product of $Z/\left( Z+1\right) $ by a strictly
decreasing function of $Z$ reaching $0$ for $Z=1.\;$Hence $\omega =+1,$ and $%
\phi ^{*}$ is then read from (\ref{statE31/phi}). The last assertion is
straightforward.
\end{proof}

Now consider $Eq_{1}$ deprived from the terms such that $2p+1\leq x\leq 2p+2$%
; we call this $Eq_{1}^{**}$. Obviously $Eq_{1}<Eq_{1}^{**}$, and from Lemma
\ref{monotphi(x>2p)} $Eq_{1}^{**}$ decreases for $0\leq \phi \leq 1$. Thus,
to prove Claim 1 it suffices to show that for any $\beta _{1}\in \left[
\beta _{1min},\beta _{1max}\right] ,\;\;Eq_{1}^{**}$ is negative at $\phi
^{*}$. Call the corresponding values of $U$ and $V$, as functions of $\beta
_{1},$ $U^{*}$ and $V^{*}$ respectively. In a moment, we will show that $%
U^{*}/V^{*}$ and $V^{*}$ behave like their unstarred, fixed-$\phi $
counterparts, i.e., the first increases and the second decreases in $\beta
_{1}$. Then, an upper bound for $Eq_{1}^{**}$ in some interval $\left[ \beta
_{1L},\beta _{1H}\right] $ is given by $M\left[ \beta _{1L},\beta
_{1H}\right] $ defined as:
\[
\widetilde{K}-\lambda \phi ^{*}\left[ \beta _{1H}\right] -\sum_{5\leq
2p+3\leq x\leq x_{max}}\widetilde{\kappa }_{x,p}\left( x-2p\right) \left[ 1-%
\frac{1}{1+\left( \frac{U^{*}}{V^{*}}\right) \left[ \beta _{1L}\right]
^{x-2p}\left( 1-\left( \frac{1}{V^{*}\left[ \beta _{1H}\right] }\right)
^{p}\right) }\right] .
\]
Straightforward numerical calculation yields (still, of course, for $%
x_{max}=56$)
\begin{eqnarray*}
M\left[ .33,.39\right] &<&-.051,\ \ \ \ M\left[ .39,.428\right] <-.051, \\
M\left[ .428,.468\right] &<&-.062,\ \ \ \ \ \ \ \ M\left[ .468,.529\right]
<-.055,
\end{eqnarray*}
establishing our final point. Now as promised:

\begin{lemma}
$V^{*}$ is a decreasing, $U^{*}/V^{*}$ an increasing function of $\beta
_{1}\in \left[ 0,1\right] $.
\end{lemma}

\begin{proof}
Set $A=9-\sqrt{3}\sqrt{3+4\beta _{1}+4\beta _{1}^{2}}$, $B=4-2\beta _{1}$,
so that $3\leq A\leq 6,$ $2\leq B\leq 4,$ and
\[
V^{*}=\frac{4}{3}\frac{A^{2}}{\left( A-B\right) \left( A+3B\right) },\ \ \ \
\ \ \ \ \ \ \frac{U^{*}}{V^{*}}=\frac{3}{8}\frac{\left( A-B\right) ^{2}}{%
\left( 1-\beta _{1}\right) \left( 3B-A\right) }\left[ 3\left( 1+2\beta
_{1}\right) +\left( 9-A\right) \right] .
\]
We take derivatives w.r.t. $\beta _{1},$ noting that $A^{\prime }=-6\left(
1+2\beta _{1}\right) /\left( 9-A\right) $. First, $V^{*\prime }$ has the
sign of $AB^{\prime }-A^{\prime }B=2/\left( 9-A\right) \;\left[ 30\;\beta
_{1}+21-9\sqrt{3}\sqrt{3+4\beta _{1}+4\beta _{1}^{2}}\right] <0.$ As for $%
U^{*}/V^{*}$, the factor in square brackets on the right is increasing, so
it suffices to show that each of $\left( A-B\right) /\left( 1-\beta
_{1}\right) $ and $\left( A-B\right) /\left( 3B-A\right) $ increases too.
The derivative of the latter has the sign of $A^{\prime }B-AB^{\prime }$,
positive as we have just seen; the derivative of the former is $\left[
A^{\prime }\left( 1-\beta _{1}\right) +A-2\right] /\left( 1-\beta
_{1}\right) ^{2}$, so has the sign of $-6\left( 1+2\beta _{1}\right) \left(
1-\beta _{1}\right) +\left( A-2\right) \left( 9-A\right) =7\sqrt{3}\sqrt{%
3+4\beta _{1}+4\beta _{1}^{2}}-15-18\;\beta _{1}>0.$
\end{proof}

This ends the proof of Proposition \ref{sepmon}. Although it can be made
slightly more precise and then used to show the existence, uniqueness, and
globally decreasing character of the implicit functions defined by $Eq_{1}$
and $Eq_{2},$ we will not do so, since we do not need to.\ We could actually
remove the word `strict' and still proceed to the final subsection.

\subsection{Root localization.}

Using Proposition \ref{sepmon}, we shall show that \textit{any }(feasible)
common root $\left( \phi ^{*},\beta _{1}^{*}\right) $ to $Eq_{1}$ and $%
Eq_{2} $ must lie in a small rectangle $\mathcal{R=}\left[ \phi _{-},\phi
_{+}\right] \times \left[ \beta _{1-},\beta _{1+}\right] $ and that on the
whole of $\mathcal{R}$, our bound (\ref{practicalbound}) for the expectation
is strictly less than $1$. Since we already know the (global) maximizer for (%
\ref{tfb1}) to exist and to necessarily give rise to such a common root for
which, besides, (\ref{practicalbound}) will be valid, this will show our
chosen value of $c$, $4.506$, to be above the threshold without even the
need for a \textit{direct} proof of either existence or uniqueness of such a
$\left( \phi ^{*},\beta _{1}^{*}\right) .$

We determine $\mathcal{R}$ explicitly, together with four numerical
sequences which witness to the fact that no solution can lie outside $%
\mathcal{R}$, owing to Corollary \ref{rectexclusion} below. This amounts, in
a rigorous presentation, to a very elementary numerical procedure which
starts at a corner of the rectangle $\left[ \phi _{min},\phi _{max}\right]
\times \left[ \beta _{1min},\beta _{1max}\right] $ containing $\mathcal{D}%
_{\phi ,\beta _{1}},$ and spirals its way towards a solution.

From Proposition \ref{sepmon} follows

\begin{lemma}
\label{signpreservn} Let $Eq$ be either $Eq_{1}$ or $Eq_{2}$, $A=\left( \phi
_{A},\beta _{1,A}\right) $, $B=\left( \phi _{B},\beta _{1,B}\right) $.
\textit{(i) }If $Eq\left( A\right) >0$ and $B\leq A$ (meaning $\phi _{B}\leq
\phi _{A}$ and $\beta _{1,B}\leq \beta _{1,A}$), then $Eq\left( B\right) >0$%
; \textit{(ii) }If $Eq\left( A\right) <0$ and $A\leq B,$ then $Eq\left(
B\right) <0$.
\end{lemma}

\begin{proof}
Do it in two steps, changing one coordinate at a time; for \textit{(i)}, use
monotony; for \textit{(ii), }use monotony if $Eq=Eq_{2}$, and if $Eq=Eq_{1}$
use the fact that if $Eq_{1}$ is negative, then it stays so if a single
coordinate is increased.
\end{proof}

This in turn implies

\begin{proposition}
\label{bandexclusion} Let $A=\left( \phi _{0},\beta _{1,A}\right) $ and $%
B=\left( \phi _{0},\beta _{1,B}\right) $ have the same $\phi $-coordinate,
while $C=\left( \phi _{C},\beta _{1,0}\right) $ and $D=\left( \phi
_{D},\beta _{1,0}\right) $ have the same $\beta _{1}$-coordinate.\newline
\textit{(i) }If $\beta _{1,B}<\beta _{1,A}$, $Eq_{1}\left( A\right) >0$ and $%
Eq_{2}\left( B\right) <0$, then the closed rectangle $\left[ \phi
_{min},\phi _{max}\right] \times \left[ \beta _{1,B},\beta _{1,A}\right] $
contains no common root to $Eq_{1}$ and $Eq_{2}$.\newline
\textit{(ii) }If $\beta _{1,A}<\beta _{1,B}$, $Eq_{1}\left( A\right) <0$ and
$Eq_{2}\left( B\right) >0$, then the closed rectangle $\left[ \phi
_{min},\phi _{max}\right] \times \left[ \beta _{1,A},\beta _{1,B}\right] $
contains no common root to $Eq_{1}$ and $Eq_{2}$.\newline
\textit{(iii) }If $\phi _{C}<\phi _{D}$, $Eq_{2}\left( C\right) <0$ and $%
Eq_{1}\left( D\right) >0$, then the closed rectangle $\left[ \phi _{C},\phi
_{D}\right] \times \left[ \beta _{1min},\beta _{1max}\right] $ contains no
common root to $Eq_{1}$ and $Eq_{2}$.\newline
\textit{(iv) }If $\phi _{D}<\phi _{C}$, $Eq_{2}\left( C\right) >0$ and $%
Eq_{1}\left( D\right) <0$, then the closed rectangle $\left[ \phi _{D},\phi
_{C}\right] \times \left[ \beta _{1min},\beta _{1max}\right] $ contains no
common root to $Eq_{1}$ and $Eq_{2}$.\newline
\end{proposition}

\begin{proof}
\textit{(i) }Let $P=\left( \phi ,\beta _{1}\right) $ be an arbitrary point
of the rectangle. We show that if $\phi \leq \phi _{0}$ then $P$ is not a
solution of $Eq_{1},$ while if $\phi \geq \phi _{0}$, $P$ fails to satisfy $%
Eq_{2}$. Indeed, in the former case we have $P\leq A$, so we use \textit{(i)
}of Lemma \ref{signpreservn} with $Eq=Eq_{1}$; in the latter, $B\leq P$, so
we apply \textit{(ii)} of the same lemma\textit{\ }with $Eq=Eq_{2}$.\newline
\textit{(ii),} \textit{(iii)} and \textit{(iv) }Very similar (or actually
the same up to notation).
\end{proof}

As an obvious corollary, we obtain the final link leading to the main result
of this paper:

\begin{corollary}
\label{rectexclusion} Let four finite sequences $\phi _{0}^{-}<\phi
_{1}^{-}<...<\phi _{K}^{-}$, $\beta _{1,0}^{+}>\beta _{1,0}^{+}>...>\beta
_{1,K}^{+}$, $\phi _{0}^{+}>\phi _{1}^{+}>...>\phi _{L}^{+}$, $\beta
_{1,0}^{-}<\beta _{1,1}^{-}<...<\beta _{1,L}$ with $\phi _{0}^{-}=\phi _{min}
$, $\beta _{1,0}^{+}=\beta _{1max}$, $\phi _{0}^{+}=\phi _{max}$ and $\beta
_{1,0}^{-}=\beta _{1min}$ verify:
\begin{eqnarray*}
Eq_{1}\left( \phi _{i}^{-},\beta _{1,i}^{+}\right)  &>&0,\ \ 0\leq i\leq K,\
\ \ \ \ \ \ \ \ Eq_{2}\left( \phi _{i}^{-},\beta _{1,i+1}^{+}\right) <0,\ \
0\leq i\leq K-1, \\
Eq_{1}\left( \phi _{j}^{+},\beta _{1,j}^{-}\right)  &<&0,\ \ 0\leq j\leq L,\
\ \ \ \ \ \ \ \ Eq_{2}\left( \phi _{j}^{+},\beta _{1,j+1}^{-}\right) >0,\ \
0\leq j\leq L-1.
\end{eqnarray*}
Then no feasible common solution to $Eq_{1}$ and $Eq_{2}$ can lie outside
the rectangle $\left] \phi _{K}^{-},\phi _{L}^{+}\right[ \times \left] \beta
_{1,L}^{-},\beta _{1,K}^{+}\right[ $.
\end{corollary}

\begin{proof}
Successive applications of Proposition \ref{bandexclusion} \textit{(i)} with
$A=\left( \phi _{i}^{-},\beta _{1,i}^{+}\right) $ and $B=\left( \phi
_{i}^{-},\beta _{1,i+1}^{+}\right) $ exclude the band $\left[ \phi
_{min},\phi _{max}\right] \times \left[ \beta _{1,K+1}^{+},\beta
_{1max}\right] $.\ We similarly exclude $\left[ \phi _{min},\phi
_{max}\right] \times \left[ \beta _{1min},\beta _{1,L+1}^{-}\right] $, $%
\left[ \phi _{min},\phi _{K}^{-}\right] \times \left[ \beta _{1min},\beta
_{1max}\right] $, and $\left[ \phi _{L}^{+},\phi _{max}\right] \times \left[
\beta _{1min},\beta _{1max}\right] $.
\end{proof}

All that remains to do is explicitly to give our four sequences as above,
and to check that the hypotheses of the corollary obtain and that the bound (%
\ref{practicalbound}) is uniformly strictly less than one on the whole
rectangle $\mathcal{R}=\left[ \phi _{K}^{-},\phi _{L}^{+}\right] \times
\left[ \beta _{1,L}^{-},\beta _{1,K}^{+}\right] $ (the bound being
independent of sufficiently large $n$).

We first compute the product $G_{1}\left( \varepsilon ,x_{max}\right)
G_{2}\left( \varepsilon ,x_{max}\right) \exp \left( 2\varepsilon D/e\right) $
appearing in (\ref{practicalbound}), to be less than $1+10^{-7.}$.

We then determine sequences $\phi _{i}^{-}$ and $\beta _{1,i}^{+}$ as above,
with $K=62$, and sequences $\phi _{i}^{+}$ and $\beta _{1,i}^{-}$ with $L=52$
satisfying the requirements of Corollary \ref{rectexclusion}, and such that $%
\mathcal{R}=\left[ \phi _{K}^{-},\phi _{L}^{+}\right] \times \left[ \beta
_{1,L}^{-},\beta _{1,K}^{+}\right] =\left[ 0.56383217,0.56383249\right]
\times \left[ 0.44651403,0.44651478\right] $.\ Taking into account the
monotony properties of $U,V$, and of the functions $x\mapsto x^{x}$, $%
x\mapsto x^{y}$ and $x\mapsto y^{x}$, an upper bound for the right-hand side
of (\ref{practicalbound}) throughout $\mathcal{R}$ is seen to be the product
of $\left( 6n^{3}\right) ^{1/n}$ by
\begin{eqnarray*}
&&\left( 1+10^{-7}\right) \times 3^{c}\left( \frac{\lambda }{6e}\right)
^{\lambda }\frac{2^{\sum_{x>2p}\widetilde{\kappa }_{x,p}}}{%
\prod\limits_{0\leq 2p\leq x\leq x_{max}}\left[ p!\left( x-p\right) !%
\widetilde{\kappa }_{x,p}\right] ^{\widetilde{\kappa }_{x,p}}}\times \\
&&\frac{\breve{U}^{\left( \lambda \phi _{L}^{+}-\widetilde{K}\right) }}{%
\left( \breve{V}-1\right) ^{\beta _{1,K+1}^{+}c}}\prod\limits_{0\leq 2p\leq
x\leq x_{max}}\left[ \widehat{V}^{x-p}+\widehat{U}^{x-2p}\left( \widehat{V}%
^{p}-1\right) \right] ^{\widetilde{\kappa }_{x,p}}\times \\
&&\left[ \frac{\left( 3\phi _{L}^{+}-\beta _{1,L}^{-}\right) ^{3\phi
_{L}^{+}-\beta _{1,L+1}^{-}}\left[ 3\left( 1-\phi _{K}^{-}\right) \right]
^{3\left( 1-\phi _{K}^{-}\right) }}{\breve{\beta}_{2}^{\breve{\beta}%
_{2}}\left( 3\breve{\beta}_{3}\right) ^{\breve{\beta}_{3}}}\right] ^{c},
\end{eqnarray*}
where $\widehat{U}=U\left( \phi _{L}^{+},\beta _{1,K}^{+}\right) $, $%
\widehat{V}=V\left( \phi _{K}^{-},\beta _{1,L}^{-}\right) $, $\breve{U}%
=U\left( \phi _{K}^{-},\beta _{1,L}^{-}\right) $, $\breve{V}=V\left( \phi
_{L}^{+},\beta _{1,K}^{+}\right) $, $\breve{\beta}_{2}=3\left( 1-\phi
_{L}^{+}\right) -2\beta _{1,K}^{+}$, and $\breve{\beta}_{3}=\beta
_{1,L}^{-}-2+3\phi _{K}^{-}$. The product of this bound by $2^{\rho
+\varepsilon \Delta }<1+10^{-14}$ is computed to be $<0.9999885.$ So, for $%
c=4.506$, $x_{max}=56$, and $\varepsilon =10^{-15}$, the product $2^{\left(
\rho +\varepsilon \Delta \right) n}\mathbf{E}\left( X_{n,\varepsilon
,x_{max},c}\right) $ is less than $6n^{3}\;0.9999885^{n}$, and we conclude
using Proposition \ref{domtyp} and the decreasing character of $\mathbf{Pr}%
_{n,c}\left( SAT\right) $ as a function of $c$.

\hspace*{0.5cm}\textbf{Acknowledgment. }We are especially grateful to Jean
Bretagnolle for his generous help.

% POUR WINEDIT ENLEVER LA MISE EN COMMENTAIRE CI-DESSOUS JUSQU'A FIN
\appendix
%\newpage
\begin{center}
\section{\hspace{-4.3cm}Appendix}
\end{center}
% POUR WINEDIT FIN DE ENLEVER LA MISE EN COMMENTAIRE CI-DESSUS

% POUR WINEDIT MISE EN COMMENTAIRE DES 2 COMMANDES CI-DESSOUS JUSQU'A FIN
%\appendix

%\section{Appendix}

% POUR WINEDIT FIN DE MISE EN COMMENTAIRE DES 2 COMMANDES CI-DESSUS

\textbf{Proof} \textbf{of Lemma \ref{typstruct}} (As stated, what we prove
is actually stronger.) In the ordered-clauses model, if the number of
occurrences of variable $i$ is $K_{i},$ the random vector $\left(
K_{i}\right) _{1\leq i\leq n}$ follows a multinomial law of parameters $%
\lambda n,p_{1}=...=p_{n}=1/n,$ where $\lambda =3c.$ Also, the number of
positive occurrences of variable $i$ is modelled by the r.v. $S_{i}=\Sigma
_{j=1}^{K_{i}}X_{i,j},$ with $X_{i,j}$ i.i.d. $B\left( 1,1/2\right) $ coin
tosses, constructed to be independent of the multinomial vector (cf. Th.
2.19 of \cite{Kal98}). For $x\in \Bbb{N}$, $S_{i,x}=\sum_{j=1}^{x}X_{i,j}$
has a binomial $B\left( x,1/2\right) $ distribution:
\begin{equation}
\mathbf{\Pr }\left( S_{i}=p\left| K_{i}=x\right. \right) =\mathbf{\Pr }%
\left( S_{i,x}=p\left| K_{i}=x\right. \right) =\mathbf{\Pr }\left(
S_{i,x}=p\right) =\frac{1}{2^{x}}{{{{{{{{{{{{{{{{{\binom{x}{p}}}}}}}}}}}}}}}}%
}}  \label{binoml1}
\end{equation}
The number of variables having $x$ occurrences is $N_{x}=\Sigma _{i=1}^{n}%
\mathbf{1}_{\left\{ K_{i}=x\right\} }$, while those having $x$ occurrences
out of which $p$ are positive is $W_{x,p}=\sum_{i=1}^{n}\mathbf{1}_{\left\{
K_{i}=x,S_{i}=p\right\} }$.\newline
We use the large deviation property of binomial r.v.s in the following form:%
\newline
Define $h\left( q,t\right) =\left( q+t\right) \mathrm{Log}\left(
1+t/q\right) +\left( 1-q-t\right) \mathrm{Log}\left( 1-t/\left( 1-q\right)
\right) $ if $t\leq 1-q,\;\;+\infty $ otherwise; and $c\left( q,t\right)
=\min \left\{ h\left( q,t\right) ,h\left( 1-q,t\right) \right\} .$ Let $Y$\
be the sum of $n$ independent indicator variables with common expectation $q$%
; then for any $\varepsilon >0,$%
\begin{equation}
\mathbf{\Pr }\left( \left| \frac{Y}{n}-q\right| \geq \varepsilon \right)
\leq 2e^{-c\left( q,\varepsilon \right) n}.  \label{binLD}
\end{equation}
We also use the fact that a Poisson r.v. with integral mean $\mu $ cannot
have too small a probability of equalling $\mu $: there is, as can be seen
from a variant of Stirling's formula, an absolute constant $C_{0}>0$ such
that if $Z$ is Poisson with integer parameter $\mu \geq 1$, then
\begin{equation}
\mathbf{\Pr }\left( Z=\mu \right) \geq \frac{C_{0}}{\sqrt{\mu }}.
\label{Poisslowbnd}
\end{equation}

Recall that the Poisson probability mass function, $e^{-\lambda }\lambda
^{x}/x!$, is denoted by $p\left( x,\lambda \right) $.

Now consider a Poisson r.v. $M$ with mean $\lambda n$, and construct (e.g.,
Lemma 5.9 in \cite{Kal98}) a random vector $\left( L_{i}\right) ,1\leq i\leq
n$ that is multinomially distributed conditionally on $M$, i.e.:
\[
\mathbf{\Pr }\left( \left( L_{i}\right) =\left( l_{i}\right) |M=m^{\prime
}\right) =\left(
\begin{array}{c}
m^{\prime } \\
l_{1}\;l_{2}\;...\;l_{n}
\end{array}
\right) n^{-m^{\prime }}.
\]
Probabilities and expectations in the Poissonized model will be subscripted
with a $\lambda $. In particular,
\begin{equation}
\mathbf{Pr}_{\lambda }\left( \left( L_{i}\right) =\left( l_{i}\right)
|M=\lambda n\right) =\mathbf{\Pr }\left( \left( K_{i}\right) =\left(
l_{i}\right) \right) .  \label{recoverKi}
\end{equation}
The law of the vector $\left( L_{i}\right) $ is obtained by deconditioning,
giving a sum with just one nonvanishing term:
\begin{eqnarray*}
\mathbf{Pr}_{\lambda }\left( \left( L_{i}\right) =\left( l_{i}\right)
\right)  &=&\sum_{m^{\prime }}\mathbf{Pr}_{\lambda }\left( \left(
L_{i}\right) =\left( l_{i}\right) |M=m^{\prime }\right) \mathbf{Pr}_{\lambda
}\left( M=m^{\prime }\right)  \\
&=&\frac{\left( \sum l_{i}\right) !}{\prod l_{i}!}n^{-\sum l_{i}}\times
p\left( \sum l_{i},\lambda n\right) =\prod_{i=1}^{n}e^{-\lambda }\frac{%
\lambda ^{l_{i}}}{l_{i}!}.
\end{eqnarray*}
So (summing w.r.t. all coordinates but one), the $L_{i}$ are independent,
each being Poisson with mean $\lambda $. We let $X_{i,j}^{\prime }$ be
i.i.d. coin tosses in the Poissonized model \textit{which are also
(completely) independent of the vector }$\left( L_{i},M\right) ,$ so that in
this model, the `number of occurrences of variable $i$' is $S_{i}^{\prime
}=\sum_{j=1}^{L_{i}}X_{i,j}^{\prime }$.\ We also consider, for $x\in \Bbb{N}$%
, $S_{i,x}^{\prime }=\sum_{j=1}^{x}X_{i,j}^{\prime }$, which has a binomial $%
B\left( x,1/2\right) $ distribution; on account of our independence
hypotheses
\begin{equation}
\mathbf{Pr}_{\lambda }\left( S_{i}^{\prime }=p|L_{i}=x\right) =\mathbf{Pr}%
_{\lambda }\left( S_{i,x}^{\prime }=p|L_{i}=x\right) =\mathbf{Pr}_{\lambda
}\left( S_{i,x}^{\prime }=p\right) =\frac{1}{2^{x}}{{{{{{{{{{{{{{{{{\binom{x%
}{p}}}}}}}}}}}}}}}}}}.  \label{binoml2}
\end{equation}
In terms of the indicators $U_{i}\left( x\right) =\mathbf{1}_{\left\{
L_{i}=x\right\} }$, $V_{i}\left( p\right) =\mathbf{1}_{\left\{ S_{i}^{\prime
}=p\right\} }$, and $W_{i,x,p}^{\prime }=\mathbf{1}_{\left\{
L_{i}=x,S_{i}^{\prime }=p\right\} }=U_{i}\left( x\right) V_{i}\left(
p\right) $, the `number of variables with $x$ occurrences, $p$ among them
positive' is $W_{x,p}^{\prime }=\sum_{i=1}^{n}W_{i,x,p}^{\prime }$. We will
need the following lemma, to be proved later.

\begin{lemma}
\label{depoiss} In the setup just defined, the law of $W_{x,p}^{\prime }$,
conditional on $M=\lambda n$, is the same as the law of $W_{x,p}$.
\end{lemma}

Clearly then, by (\ref{binoml2}):\newline
\[
\mathbf{E}_{\lambda }W_{i,x,p}^{\prime }=\mathbf{Pr}_{\lambda }\left(
S_{i}^{\prime }=p|L_{i}=x\right) \mathbf{Pr}_{\lambda }\left( L_{i}=x\right)
=\frac{1}{2^{x}}{{{{{{{{{{{{{{{{{\binom{x }{p}}}}}}}}}}}}}}}}}}p\left(
x,\lambda \right) \equiv \kappa _{x,p}.
\]
\newline
By (\ref{binLD}) we have:
\[
\mathbf{Pr}_{\lambda }\left( \left| \frac{W_{x,p}^{\prime }}{n}-\kappa
_{x,p}\right| \geq \varepsilon \right) \leq 2e^{-c\left( \varepsilon ,\kappa
_{x,p}\right) n}.
\]
We now \textit{depoissonize}, i.e. we decompose w.r.t. the values of $M$:
\begin{eqnarray*}
2e^{-c\left( \varepsilon ,\kappa _{x,p}\right) n} &\geq &\sum_{m^{\prime }}%
\mathbf{Pr}_{\lambda }\left( \left| \frac{W_{x,p}^{\prime }}{n}-\kappa
_{x,p}\right| \geq \varepsilon \;\left| M=m^{\prime }\right. \right) \mathbf{%
Pr}_{\lambda }\left( M=m^{\prime }\right) \\
&\geq &\mathbf{Pr}_{\lambda }\left( \left| \frac{W_{x,p}^{\prime }}{n}%
-\kappa _{x,p}\right| \geq \varepsilon \;\left| M=\lambda n\right. \right)
\mathbf{Pr}_{\lambda }\left( M=\lambda n\right) .
\end{eqnarray*}
By Lemma (\ref{depoiss}) and (\ref{Poisslowbnd}), this implies
\[
\mathbf{\Pr }\left( \left| \frac{W_{x,p}}{n}-\kappa _{x,p}\right| \geq
\varepsilon \right) \leq \frac{2}{C_{0}}\sqrt{\lambda n}e^{-c\left(
\varepsilon ,\kappa _{x,p}\right) n},
\]
which is stronger than Lemma (\ref{typstruct}).

\textbf{Proof of lemma \ref{depoiss}}. The law of a random vector determines
that of the sum of its components, and the same holds for the conditional
laws relative to some event (e.g., \cite{Loe77}, p. 218, end of \S\ 14).%
\newline
So, it is sufficient to show that the law of the random vector $\left(
W_{ixp}^{\prime }\right) _{1\leq i\leq n}$, conditional on $M=\lambda n$, is
the same as the law of $\left( W_{i,x,p}\right) _{1\leq i\leq n}$, where $%
W_{i,x,p}=\mathbf{1}_{\left\{ K_{i}=x,S_{i}=p\right\} }=\mathbf{1}_{\left\{
K_{i}=x\right\} }\mathbf{1}_{\left\{ S_{i}=p\right\} }$; and this, in turn,
will follow if we show that the conditional law of the $2n$-dimensional
random vector $\left( L_{i},S_{i}^{\prime }\right) _{1\leq i\leq n}$ is the
same as the law of $\left( K_{i},S_{i}\right) _{1\leq i\leq n}$. Now, for
any $\Bbb{N}^{n}$-valued vectors $\left( l_{i}\right) _{1\leq i\leq n}$ and $%
\left( p_{i}\right) _{1\leq i\leq n}$,
\[
\mathbf{Pr}_{\lambda }\left( \left( L_{i}=l_{i},S_{i}^{\prime }=p_{i}\right)
\left| M=\lambda n\right. \right) =\mathbf{Pr}_{\lambda }\left( \left(
L_{i}=l_{i},S_{i,l_{i}}^{\prime }=p_{i}\right) \left| M=\lambda n\right.
\right)
\]
Here the event $A=\bigcap_{1\leq i\leq n}\left\{ S_{i,l_{i}}^{\prime
}=p_{i}\right\} $ is independent of the conjunction $B\cap C$, with $%
B=\bigcap_{1\leq i\leq n}\left\{ L_{i}=l_{i}\right\} $ and $C=\left\{
M=\lambda n\right\} $, so $\mathbf{Pr}_{\lambda }\left( A\left| B\cap
C\right. \right) =\mathbf{Pr}_{\lambda }\left( A\right) $. Applying the
generally-valid
\[
\mathbf{P}\left( A\cap B\left| C\right. \right) =\mathbf{P}\left( A\left|
B\cap C\right. \right) \mathbf{P}\left( B\left| C\right. \right) ,
\]
and using (\ref{recoverKi}), we see that
\begin{equation}
\mathbf{Pr}_{\lambda }\left( \left( L_{i}=l_{i},S_{i}^{\prime }=p_{i}\right)
\left| M=\lambda n\right. \right) =\mathbf{Pr}_{\lambda }\left( \left(
S_{i,l_{i}}^{\prime }=p_{i}\right) \right) \mathbf{Pr}_{\lambda }\left(
\left( K_{i}=l_{i}\right) \right) .  \label{LS'law}
\end{equation}
Although the $K_{i}$ are \textit{not} independent, our setup does ensure
that the events $\bigcap_{1\leq i\leq n}\left\{ S_{i,l_{i}}=p_{i}\right\} $
and $\bigcap_{1\leq i\leq n}\left\{ K_{i}=l_{i}\right\} $ are independent,
so
\begin{equation}
\mathbf{\Pr }\left( \left( K_{i}=l_{i},S_{i,l_{i}}=p_{i}\right) \right) =%
\mathbf{\Pr }\left( \left( S_{i,l_{i}}=p_{i}\right) \right) \mathbf{\Pr }%
\left( \left( K_{i}=l_{i}\right) \right) .  \label{KSlaw}
\end{equation}
But, by (\ref{binoml2}) and (\ref{binoml1}), the first factors on the right
in (\ref{LS'law}) and (\ref{KSlaw}) are both equal to $%
\prod_{i=1}^{n}2^{-l_{i}}{{{{{{{{{{{{{{{{{\binom{l_{i} }{p_{i}}}}.}}}}}}}}}}}%
}}}}$ {\hfill }${\blacksquare }$

\end{document}